\documentclass{aa}

\usepackage{graphicx}
\usepackage{amssymb,amsmath}
\usepackage{natbib}
\usepackage{txfonts}
\usepackage{pifont}
\usepackage{multirow}
\usepackage{hhline}

\def\Msun{ M_\odot}

\def\Lsun{ L_\odot}

\def\Loo{1~\Lsun}
\def\Lot{10^{-2}~\Lsun}
\def\Lof{10^{-4}~\Lsun}

\def\Mp{M_{\rm p}}
\def\Rp{R_{\rm p}}

\def\Ms{M_{\star}}

\def\Ls{L_{\star}}

\def\Mearth{ M_\oplus}
\def\Rearth{ R_\oplus}

\def\ss{\sigma_{\star}}

\def\hice{h_{ice}}

\def\m{\mathrm{m}}
\def\km{\mathrm{km}}
\def\sec{\mathrm{s}}
\def\kg{\mathrm{kg}}

\def\deg{^\circ}
\def\K{\mathrm{K}}
\def\J{\mathrm{J}}
\def\inertia{\J~\sec^{-1/2}~\m^{-2}~\K^{-1}}

\usepackage[normalem]{ulem}
\usepackage{color}
\definecolor{blue}{RGB}{0,0,255}
\definecolor{red}{RGB}{255,0,0}
\definecolor{green}{RGB}{0,200,0}
\definecolor{black}{RGB}{0,0,0}

\begin{document}

\title{Habitability of planets on eccentric orbits: the limits of the mean flux approximation.} 
\titlerunning{Habitability of planets on eccentric orbits}

   \subtitle{ }

   \author{Emeline Bolmont \inst{1}
          \and Anne-Sophie Libert \inst{1}
          \and Jeremy Leconte \inst{2,3,4}
          \and Franck Selsis \inst{5,6} 
                }
\authorrunning{Emeline Bolmont, et al.}

\institute{NaXys, Department of Mathematics, University of Namur, 8 Rempart de la Vierge, 5000 Namur, Belgium
\and Canadian Institute for Theoretical Astrophysics, 60st St George Street, University of Toronto, Toronto, ON, M5S3H8, Canada
\and Banting Fellow
\and Center for Planetary Sciences, Department of Physical \& Environmental Sciences, University of Toronto Scarborough, Toronto, ON, M1C 1A4, Canada
\and Univ. Bordeaux, LAB, UMR 5804, F-33270, Floirac, France
\and CNRS, LAB, UMR 5804, F-33270, Floirac, France
}

\date{Received xxx ; accepted xxx}

\abstract{Contrary to Earth, which has a small orbital eccentricity, some exoplanets discovered in the insolation habitable zone (HZ) have high orbital eccentricities (e.g., up to an eccentricity of $\sim0.97$ for HD~20782~b). 
This raises the question of the capacity of these planets to host surface liquid water.

In order to assess the habitability of an eccentric planet, the mean flux approximation is often used. 
It states that a planet on an eccentric orbit is called habitable if it receives on average a flux compatible with the presence of surface liquid water.
However, as the planets do experience important insolation variations over one orbit and even spend some time outside the HZ for high eccentricities, the question of their habitability might not be as straightforward. 

We performed a set of simulations using the Global Climate Model LMDz, exploring the limits of the mean flux approximation when varying the luminosity of the host star and the eccentricity of the planet.
We computed the climate of tidally locked ocean covered planets with orbital eccentricity from 0 to 0.9 receiving a mean flux equal to Earth's, around stars of luminosity ranging from $\Loo$ to $\Lof$. 

Using here a definition of habitability based on the presence of surface liquid water, we find that most of the planets considered can sustain surface liquid water on the dayside with an ice cap on the nightside.
However, for high eccentricity and high luminosity, planets cannot sustain surface liquid water during the whole orbital period. 
They completely freeze at apoastron and when approaching periastron an ocean appears around the substellar point.

We conclude that the higher the eccentricity and the higher the luminosity of the star, the less reliable the mean flux approximation.}


\maketitle

\section{Introduction}
\label{intro}

The majority of the planets found in the insolation habitable zone (HZ, zone in which a planet could sustain surface liquid water, as defined by \citealt{Kasting1993}) are on eccentric orbits. 
The actual percentage depends on the definition of the inner and outer edges considered for the HZ. For instance, 
about 80\% of the planets spending some time in the conservative HZ, whose the inner edge corresponds to the ``runaway greenhouse'' criterium and the outer edge to the ``maximum greenhouse'' criterium \citep[e.g.][]{Kopparapu2013, Kopparapu2014} have an eccentricity of more than 0.1\footnote{\url{http://physics.sfsu.edu/~skane/hzgallery/index.html}}.  

While most of the planets detected in the HZ are very massive planets and probably gaseous, five of them have masses below $10~\Mearth$ and eleven of them have radii smaller than $2~\Rearth$ (such as Kepler-186f with an estimated eccentricity of $\sim 0.01$, \citealt{Quintana2014}).
Among these sixteen possibly rocky planets, four of them have eccentricities higher than 0.1: GJ~832~c \citep{Bailey2009}, Kepler-62~e, Kepler-69~c \citep{Borucki2011} and GJ~667~Cc \citep{Anglada-Escude2012, RobertsonMahadevan2014}.
Table \ref{tab_obs_ecc} shows the characteristics of these four planets as well as the percentage of the orbital phase spent within the HZ for two different definitions of the inner and outer edges.

We expect more small planets to be discovered in the HZ with the future missions to increase the statistics (e.g., NGTS, TESS) and also better constrain the eccentricity (e.g., PLATO).
In any case, this discovery raises the question of the potential habitability of planets that, as GJ~832~c and GJ~667~Cc, only spend a fraction of their orbit in the HZ.

\begin{table*}[htbp]
\begin{center}
\caption{Possibly rocky observed exoplanets with an eccentricity higher than 0.1 \citep[from the Habitable Zone Gallery,][]{KaneGelino2012}. HZ$_{\rm in-out, c}$ corresponds to the inner and outer edge of the conservative HZ and HZ$_{\rm in-out, o}$ corresponds to the inner and outer edge of the optimistic HZ \citep[the inner edge corresponds to the ``recent Venus'' criterium and the outer edge to the ``early Mars'' criterium; e.g.][]{Kopparapu2013, Kopparapu2014}.
$\tau_{\rm HZ, c}$ is the percentage of the orbital phase spent within the conservative HZ and $\tau_{\rm HZ, o}$ is the percentage of the orbital phase spent within the optimistic HZ.}
\vspace{0.1cm}
\begin{tabular}{|c||c|c||c|c|c|c||c|c|c|c|}
\hline
Planets & $\Ms$ & $T_{\rm eff}$ & $\Mp$ & $\Rp$ & periastron-apoastron & eccentricity & HZ$_{\rm in-out, c}$ & HZ$_{\rm in-out, o}$ & $\tau_{\rm HZ, c}$ &  $\tau_{\rm HZ, o}$ \\
~ & ($\Msun$) & (K) & ($\Mearth$) & ($\Rearth$) & (au) & & (au) & (au) & (\%) & (\%) \\
\hline
Kepler-62~e & 0.69 & 4925 & & 1.6 & 0.37-0.49 & 0.13 & 0.46-0.84 & 0.37-0.89 & 29.9 & 100 \\
Kepler-69~c & 0.81 & 5640 & & 1.7 & 0.61-0.81 & 0.14 & 0.85-1.50 & 0.67-1.59 & 0 & 66.9 \\
GJ~832~c & 0.45 & 3500 & 5.4 & & 0.13-0.19 & 0.18 & 0.19-0.36 & 0.15-0.38 & 24.7 & 73.3 \\
GJ~667~Cc & 0.33 & 3350 & 3.8 & & 0.09-0.16 & 0.27 & 0.15-0.29 & 0.12-0.31 & 29.5 & 35.2 \\
\hline
\end{tabular} 
\label{tab_obs_ecc} 
\end{center}
\end{table*}

The influence of the orbital eccentricity of a planet on its climate has already been studied using various methods: Energy-Balanced Models (EBMs) and Global Climate Models (GCMs).
EBMs assume that the planet is in thermal equilibrium: it must radiate on average as much long-wave radiations to space as they receive short-wave radiations from the host star \citep{WilliamsKasting1997}. 
In such models, the radiative energy fluxes entering or leaving a cell are balanced by the dynamic fluxes of heat transported by winds into or away from the cell.
On the contrary GCMs consistently compute on a three-dimensional grid the circulation of the atmosphere using forms of the Navier-Stokes equations.
GCMs are therefore more computationally demanding but they are more accurate when simulating a climate.

Using a GCM, \citet{WilliamsPollard2002a} studied the influence of the eccentricity on the climate of Earth-like planets around a Sun-like star (with the correct distribution of continents and oceans, a 365~day orbit, a 24~hour day and a 23$\deg$ obliquity) and found that surface liquid water is possible even on very eccentric orbits.
Using a GCM, \citet{Linsenmeier2015} studied the influence of both obliquity and eccentricity for ocean covered planets orbiting a Sun-like star on a 365~day orbit and a 24~hour day, like Earth.
They found that planets with eccentricities higher than 0.2 can only sustain surface liquid water for a part of the year.

\citet{Spiegel2010} and \citet{Dressing2010} used EBMs to illustrate the effect of the evolution of eccentricity \citep[through pseudo-Milankovitch's cycles;][]{Milankovitch1941}. 
\citet{Spiegel2010} found that the increase of eccentricity of a planet may allow it to escape a frozen snowball state.
\citet{Dressing2010} found that increasing the eccentricity widens the parameter space in which the planet can only sustain surface liquid water for part of the year.



A major result of \citet{WilliamsPollard2002a} was that the capacity of an eccentric planet of semi-major axis $a$ and eccentricity $e$ to host surface liquid water depends on the averaged flux received over one orbit.
This averaged flux corresponds to the flux received by a planet on a circular orbit of radius $r = a(1-e^2)^{1/4}$. 
If this orbital distance is within the HZ, then the planet is assumed to belong to the HZ \citep[or to the eccentric HZ, as defined by][]{Barnes2008}. 
However this study was performed by simulating the climate of an Earth-twin planet.
The generalization of this result to the diversity of the planets discovered in the HZ is not straightforward.
While, we expect this mean-flux approximation to be adequate for planets with low eccentricities, for high eccentricities however, the climate could be drastically degraded when the planet is temporarily outside the HZ.
This would especially be an issue for planets around hot stars, for which the HZ is far from the star.
The planet could spend a long time outside of the HZ, leading to the freezing of the water reservoir at apoastron and its evaporation at periastron.

\medskip

The influence of the stellar luminosity/host star type has previously been considered for Earth-like planets on circular orbits (e.g., \citealt{Shields2013, Shields2014}). 
Furthermore, \cite{Wordsworth2011} have studied the climate of GJ~581d orbiting a red dwarf, for two different eccentricities (0 and 0.38).
But no work has studied jointly the influence of the planet's eccentricity and the stellar luminosity.

\medskip

We therefore aimed to explore here, in a systematic way, the influence of the planet's eccentricity and the star luminosity on the climate of ocean covered planets in a 1:1 spin-orbit resonance, receiving on average the same flux as Earth. 
In order to test the limits of the mean flux approximation, we performed three-dimensional GCM simulations for a wide range of configurations: we considered stars of luminosity $\Loo$, $\Lot$ and $\Lof$ and orbits of eccentricity from 0 to 0.9.
We took into account the different luminosities by scaling the orbital period of the planets. 
It means that we did not consider here the spectral dependance of the stars.
We investigated the capacity of these planets to sustain surface liquid water.

In Section \ref{definitions_HZ}, we first present our definition of habitability in terms of surface liquid water coverage.
In Section \ref{model}, we explain the set-up of our simulations and in Sections \ref{Circular} and \ref{Eccentric}, we discuss their outcome in terms of liquid water coverage.
In Section \ref{Observables}, we discuss the observability of the variability caused by eccentricity.
Finally, in Sections \ref{Discussion} and \ref{Conclusions}, we conclude this study.

\section{Liquid water coverage vs. habitability}
\label{definitions_HZ}

In this article, we do not consider that habitability is equivalent to the requirement of having a mean surface temperature higher than the freezing point of water, like in the energy balance models \citep[e.g.,][]{Williams1996Kasting} or the radiative-convective models \citep[e.g.,][]{Kasting1993}.
As in \citet{Spiegel2008}, we choose here an assessment of the habitability of a planet based on sea ice cover. 
We focus here only on the presence of surface liquid water without having to conclude about the actual potential of the planets to be appropriate environments for the apparition of life.

The planets considered in this work are water worlds (or aqua worlds), i.e. planets whose whole surface is covered with water (here treated as an infinite water source). 
Considering water worlds is especially convenient for a first study because it allows us to have a small amount of free parameters (no land/ocean distribution, land roughness, etc...).
A subset of this population is the ocean planets with a high bulk water fraction, which strongly alters their internal structure.
Ocean planets were hypothesized in the early 2000s by \citet{Kuchner2003} and \citet{Leger2004}.
They are believed to have a mass ranging from $1~\Mearth$ (small rocky planets) to $10~\Mearth$ (mini-Neptunes).
Their composition was investigated, and the depth of the ocean of a Earth-mass planet was estimated to a few hundred kilometers \citep{Sotin2007}. 
These planets could be identified providing that we would know mass and radius with enough precision \citep{Sotin2007,Selsis2007}.
Despite the lack of knowledge on their mass, some observed planets have been proposed to be ocean planets, for example Kepler-62e and -62f \citep{Kaltenegger2013}.

%


\begin{table*}[htbp]
\begin{center}
\caption{Planets' orbit characteristics and received flux for $\Ls = \Lsun$. $a$ is the semi-major axis defined in Eq. \ref{equiv_radius}. The planets being in synchronous rotation, the orbital period (P$_{\rm orb}$ columns, given in Earth's day = 24~hr) and the rotation period of the planet are equal. Peri denotes the periastron distance and apo the apoastron distance.}
\vspace{0.1cm}
\begin{tabular}{|r||r|r|r|r|r|r||}
\hhline{~------}
 \multicolumn{1}{r||}{}  & \multicolumn{6}{c||}{$\Ls = \Lsun$} \\
\hhline{-------}
 \multicolumn{1}{|c||}{ecc} & a (au) & peri (au) & apo (au) & P$_{\rm orb}$ (day) & Flux at peri (W/m$^2$) & Flux at apo (W/m$^2$) \\
\hhline{-------}
\multicolumn{1}{|c||}{0} & 1.000 & 1.00 & 1.00 & 365.5  & 1,366 & 1,366 \\
\multicolumn{1}{|c||}{0.05} & 1.001 & 0.95 & 1.05 & 365.9  & 1,517 & 1,241 \\
\multicolumn{1}{|c||}{0.1} & 1.003 & 0.90 & 1.10 & 366.9 & 1,697 & 1,136 \\
\multicolumn{1}{|c||}{0.2} & 1.011 & 0.81 & 1.21 & 371.2  & 2,128 & 954 \\
\multicolumn{1}{|c||}{0.4}  & 1.045 & 0.63 & 1.46 & 390.2  & 3,758 & 700 \\
\multicolumn{1}{|c||}{0.6} & 1.119 & 0.45 & 1.79 &  432.1  & 8,447 & 534 \\
\multicolumn{1}{|c||}{0.8} & 1.292 & 0.26 & 2.33 &  536.2  & 33,731 & 420 \\
\multicolumn{1}{|c||}{0.9} & 1.516 & 0.15 & 2.88 &  681.4  & 139,530 & 378 \\
\hhline{-------}
\end{tabular} 
\label{tab:init1} 
\end{center}
\end{table*}

\begin{table*}[htbp]
\begin{center}
\caption{Planets' orbit characteristics for $\Ls = \Lot$ and $\Ls = \Lof$. The fluxes at periastron and apoastron are the same as in Table \ref{tab:init1}.}
\vspace{0.1cm}
\begin{tabular}{r||r|r|r|r||r|r|r|r||}
\hhline{~--------}
  & \multicolumn{4}{c||}{$\Ls = \Lot$} & \multicolumn{4}{c||}{$\Ls = \Lof$}  \\
\hhline{---------}
 \multicolumn{1}{|c||}{ecc} & a (au) & peri (au) & apo (au) & P$_{\rm orb}$ (day) & a (au) & peri (au) & apo (au) & P$_{\rm orb}$ (day)\\
\hline
\multicolumn{1}{|c||}{0} & 1.000$\times 10^{-1}$ & 1.00$\times 10^{-1}$ & 1.00$\times 10^{-1}$ & 22.85  & 1.000$\times 10^{-2}$ & 1.00$\times 10^{-2}$ & 1.00$\times 10^{-2}$ &  1.967\\
\multicolumn{1}{|c||}{0.05} & 1.001$\times 10^{-1}$ & 0.95$\times 10^{-1}$ & 1.05$\times 10^{-1}$ & 22.87 & 1.001$\times 10^{-2}$ & 0.95$\times 10^{-2}$ & 1.05$\times 10^{-2}$ & 1.968 \\
\multicolumn{1}{|c||}{0.1} & 1.003$\times 10^{-1}$ & 0.90$\times 10^{-1}$ & 1.10$\times 10^{-1}$ & 22.94 & 1.003$\times 10^{-2}$ & 0.90$\times 10^{-2}$ & 1.10$\times 10^{-2}$  & 1.974 \\
\multicolumn{1}{|c||}{0.2} & 1.011$\times 10^{-1}$ & 0.81$\times 10^{-1}$ & 1.21$\times 10^{-1}$ & 23.21 & 1.011$\times 10^{-2}$ & 0.81$\times 10^{-2}$ & 1.21$\times 10^{-2}$  & 1.997 \\
\multicolumn{1}{|c||}{0.4}  & 1.045$\times 10^{-1}$ & 0.63$\times 10^{-1}$ & 1.46$\times 10^{-1}$ & 24.40 & 1.045$\times 10^{-2}$ & 0.63$\times 10^{-2}$ & 1.46$\times 10^{-2}$  & 2.099 \\
\multicolumn{1}{|c||}{0.6} & 1.119$\times 10^{-1}$ & 0.45$\times 10^{-1}$ & 1.79$\times 10^{-1}$ & 27.02 & 1.119$\times 10^{-2}$ & 0.45$\times 10^{-2}$ & 1.79$\times 10^{-2}$ & 2.325 \\
\multicolumn{1}{|c||}{0.8} & 1.292$\times 10^{-1}$ & 0.26$\times 10^{-1}$ & 2.33$\times 10^{-1}$ & 33.52  & 1.292$\times 10^{-2}$ & 0.26$\times 10^{-2}$ & 2.33$\times 10^{-2}$ & 2.885 \\
\multicolumn{1}{|c||}{0.9} & 1.516$\times 10^{-1}$ & 0.15$\times 10^{-1}$ & 2.88$\times 10^{-1}$ & 42.60  & 1.516$\times 10^{-2}$ & 0.15$\times 10^{-2}$ & 2.88$\times 10^{-2}$ & 3.666 \\
\hline
\end{tabular} 
\label{tab:init2} 
\end{center}
\end{table*}

%

\section{GCM simulations}
\label{model}

\subsection{Model parameters}

We performed the climate simulations with the LMD generic Global Climate Model (GCM) widely used for the study of extrasolar planets \citep[e.g.][]{Wordsworth2010,Wordsworth2011,Selsis2011} and the paleoclimates of Mars \citep{Wordsworth2013,Forget2013}. 
In particular, we used the 3D dynamical core of the LMDZ 3 GCM \citep{Hourdin2006}, based on a finite-difference formulation of the primitive equations of geophysical fluid dynamics. 
A spatial resolution of $64$x$48$x$30$ in longitude, latitude, and altitude was set for the simulations.

We assumed that the atmosphere is mainly composed of N$_2$, with 376~ppmv of CO$_2$, which corresponds to an Earth-like atmosphere. 
The water cycle was modeled with a variable amount of water vapor and ice. 
Ice formation (melting) was assumed to occur when the surface temperature is lower (higher) than 273 K, and temperature changes due to the latent heat of fusion were taken into account. 

We used \citet{Leconte2013a}'s computed high-resolution spectra over a range of temperatures and pressures using the HITRAN 2008 database \citep{Rothman2009}. 
We adopted the same temperature grids as in \citet{Leconte2013a} with values $T = $\{110,170,...,710\}~K and the same pressure grids with values $p = $\{10$^{-3}$,10$^{-2}$,...,10$^{5}$\}~mbar.
The water volume mixing ratio could vary between 10$^{-7}$ to 1. 
The H$_2$O absorption lines were truncated at 25~cm$^{-1}$, but the water vapor continuum was included using the CKD model \citep{Clough1989}.
As in \citet{Leconte2013a}, the opacity due to N$_2$--N$_2$ collision-induced absorption was also taken into account.

We used the same correlated-k method as in \citet{Wordsworth2011} and \citet{Leconte2013a} to produce a smaller database of spectral coefficients suitable for fast calculation of the radiative transfer in the GCM.
The model used 38 spectral bands for the thermal emission of the planet and 36 for the stellar emission.
In our water world model, we did not take into account the oceanic circulation. 
We chose an albedo of 0.07 for the surface liquid water and an albedo of 0.55 for the ice and snow.

\medskip


We considered the influence of some parameters on the outcome of our simulations: the thermal inertia $I_{oc}$ of the oceans and the maximum ice thickness $h_{ice}$ allowed in our model.
We tested the following 3 combinations: 
 \begin{equation*}
\begin{cases}
h_{ice} = 1~\m, & I_{oc} = 18000~\inertia, \\
h_{ice} = 10~\m, & I_{oc} = 18000~\inertia, \\
h_{ice} = 1~\m, & I_{oc} = 36000~\inertia.
\end{cases}
\end{equation*}
The results of Sections \ref{Circular}, \ref{Eccentric} have been obtained with the first combination, while comparison with the two other sets of values is performed in Section \ref{therm_inert_hice}.
Changing thermal inertia is a way to model the ocean mixed layer depth which responds quickly to the climatic forcing. 
This mixed layer varies on Earth with location and time.
\citet{Selsis2013} studied the effect of changing the thermal inertia of hot planets without atmospheres. 
They showed that increasing the thermal inertia of the surface of such a planet damped the amplitude of its temperature response to the eccentricity-driven insolation variations. 
They also showed that increasing thermal inertia introduced a lag of the response of the planet with respect to the insolation variations.
Imposing a maximum ice thickness allowed in the model is a way to mimic oceanic transports which limits the growth of ice layers. 
It influences the time to reach the equilibrium and the eccentricity-driven oscillations (see Section \ref{therm_inert_hice}).

\subsection{Planets and initial conditions}\label{initial_conditions_planets}

We computed the climate of water worlds, initially ice-free, receiving a mean flux equal to Earth's ($1366$~W/m$^2$) on orbits of eccentricity from 0 to 0.9, around stars of luminosity:
\begin{itemize}
\item $\Ls = \Loo$, corresponding to our Sun with an effective temperature of $\sim 5800$~K;
\item $\Ls = \Lot$, corresponding to a M-dwarf of $0.25~\Msun$ with an effective temperature of $\sim 3300$~K;
\item $\Ls = \Lof$, corresponding to a $500$~Myr old brown-dwarf of mass $0.04~\Msun$ with an effective temperature of $\sim 2600$~K.
\end{itemize}

Note that we did not take into account in this work the spectral dependance of the stars. 
For instance, we did not consider that a $10^{-4}~\Lsun$ star is much redder than a $1~\Lsun$ star. 
We took into account the different luminosities only by scaling the orbital period of the planets, as explained in the following.

\medskip

We considered here planets in a 1:1 spin-orbit resonance, regardless of their eccentricity.
However, a planet orbiting a $1~\Lsun$ star on a circular orbit and receiving a flux of $1366$~W/m$^2$ (i.e. at an orbital distance of $1$~au) will not reach a synchronous rotation state in less than the age of the universe. 
Moreover, if the planet is very eccentric, the probability that it is in synchronous rotation is low. 
The planet will more likely be either in pseudo-synchronization \citep[synchronization at periastron,][]{Hut1981}, or in spin-orbit resonance \citep{Makarov2013}.
The aim of our work was to investigate the effect of eccentricity and luminosity, so we only varied here these two parameters, keeping all others equal.
Choosing a synchronous rotation allowed us to have, for a given eccentricity, the exact same insolation evolution for planets orbiting a high luminosity star as a low luminosity star.
The obliquity of the planet was assumed to be zero.
In all cases, the simulations were run from an initial ice-free state until runaway greenhouse/glaciation occurred or steady states of thermal equilibrium were reached. 


\medskip

For the different eccentricities, we scaled the orbital period of the planet (the duration of the ``year'') to insure that the planet receives $F_\oplus = 1366$~W/m$^2$ on average. 
A planet of semi-major axis $a$ and eccentricity $e$ receives an averaged flux over one orbit of:
\begin{equation} \label{flux_ae}
F = \frac{\Ls}{4\pi a^2\sqrt{1-e^2}},
\end{equation}
where $\Ls$ is the luminosity of the star. 
In our study, we assume that the planet receives on average $F_\oplus$:
\begin{equation} \label{flux_earth}
F = F_\oplus = \frac{\Lsun}{4\pi a_\oplus^2} = 1366~\mathrm{W/m}^2,
\end{equation}
where $a_\oplus = 1~$au.
Thus, we can express the semi-major axis of the planet as a function of eccentricity $e$ and stellar luminosity $\Ls$:
\begin{equation} \label{equiv_radius}
a = \frac{a_\oplus}{(1-e^2)^{1/4}} \sqrt{\frac{\Ls}{\Lsun}}.
\end{equation}
If we increase the eccentricity of the orbit of the planet, its semi-major axis $a$ increases to insure it receives on average $F_\oplus$. 
For example, a planet around a Sun-like star with an eccentricity of 0.6 and receiving 1366~W/m$^2$ on average has a semi-major axis of 1.119~au.

Table \ref{tab:init1} shows the different values of the semi-major axis of the planets for different eccentricities, as well as the distances of periastron and apoastron, and the fluxes the planet receives at these distances for $\Ls = \Lsun$. 
Table \ref{tab:init2} shows the planets' orbit characteristics for $\Ls = \Lot$ and $\Ls = \Lof$.
Since we consider synchronous planets, we can deduce the rotation period of the planets depending on the eccentricity and the type of the star.
For a star of $\Ls=1~\Lsun$ and a planet on a circular orbit, the year is 365~days long and the planet has a slow rotation. 
If the planet is on a very eccentric orbit ($e$ = 0.9), then the semi-major axis is 1.516~au, the year is $681$~days long and the planet has an even slower rotation. 
For a star of $\Ls = 10^{-4}~\Lsun$ and a planet on a circular orbit, the semi-major is 0.01~au, the year is $\sim 2$~days long ($\sim 4$~days for $e = 0.9$) and the planet has a faster rotation.

Besides, as shown in \citet{Selsis2013}, because of the optical libration due to the 1:1 spin-orbit resonance, there is no permanent dark area on a planet with an  eccentricity higher than 0.72. 
We define here the dayside as the hemisphere which is illuminated when the planet has an eccentricity of 0. 
For this case, the substellar point is at 0$\deg$ longitude and $0\deg$ latitude, and the dayside extends to -90$\deg$ to 90$\deg$ in longitude and -90$\deg$ to 90$\deg$ in latitude. 
The nightside is the other side of the planet, i.e. from 90$\deg$ to 270$\deg$ in longitude and -90$\deg$ to 90$\deg$ in latitude.
We use this geographic definition of the dayside and nightside independently of the eccentricity of the orbit.

We will first discuss the effect of varying the star luminosity on the climate of planets on circular orbits (Section \ref{Circular}), then we will extend the discussion to planets on eccentric orbits (Section \ref{Eccentric}).


	\begin{figure*}[htbp!]
	\centering
	\includegraphics[width=\linewidth]{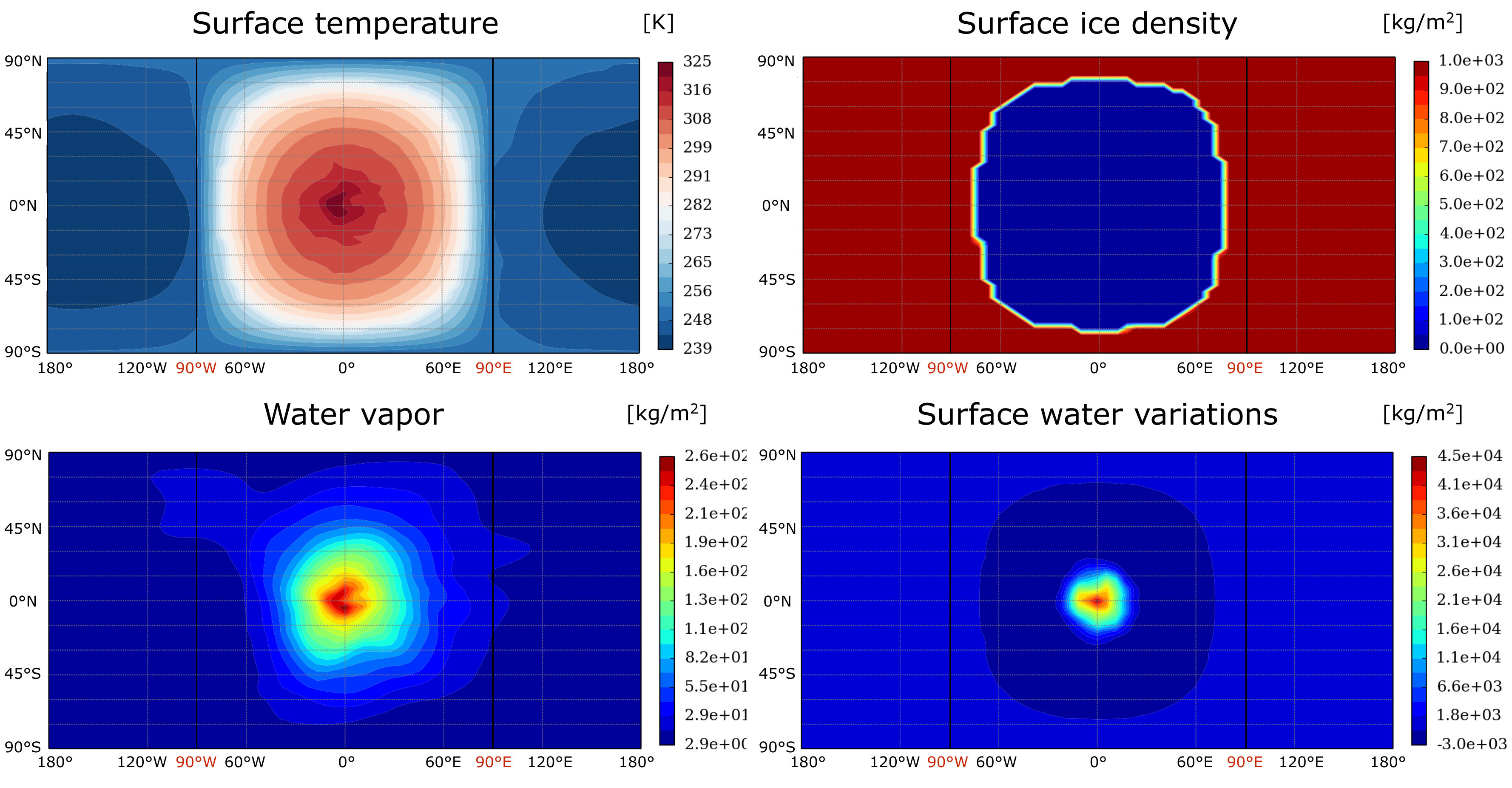}
	\caption{Maps of surface temperature, ice density, amount of water vapor integrated on a column and precipitation for a planet around a $1~\Lsun$ star on a circular orbit. The terminator (the longitudes of 90$\deg$W and 90$\deg$E) is materialized in black. For the map showing the surface water variation, a negative value means that liquid water disappears (evaporation or freezing) and a positive value means that liquid water appears (rain or melting). For the map showing the surface ice density, a density of 1000~kg/m$^2$ corresponds to a 1000~mm = 1~m thick ice layer.}
	\label{L00_ecc000_tsurf_h2o_ice_vap_col_surf}
	\end{figure*}

	\begin{figure*}[htbp!]
	\centering
	\includegraphics[width=14cm]{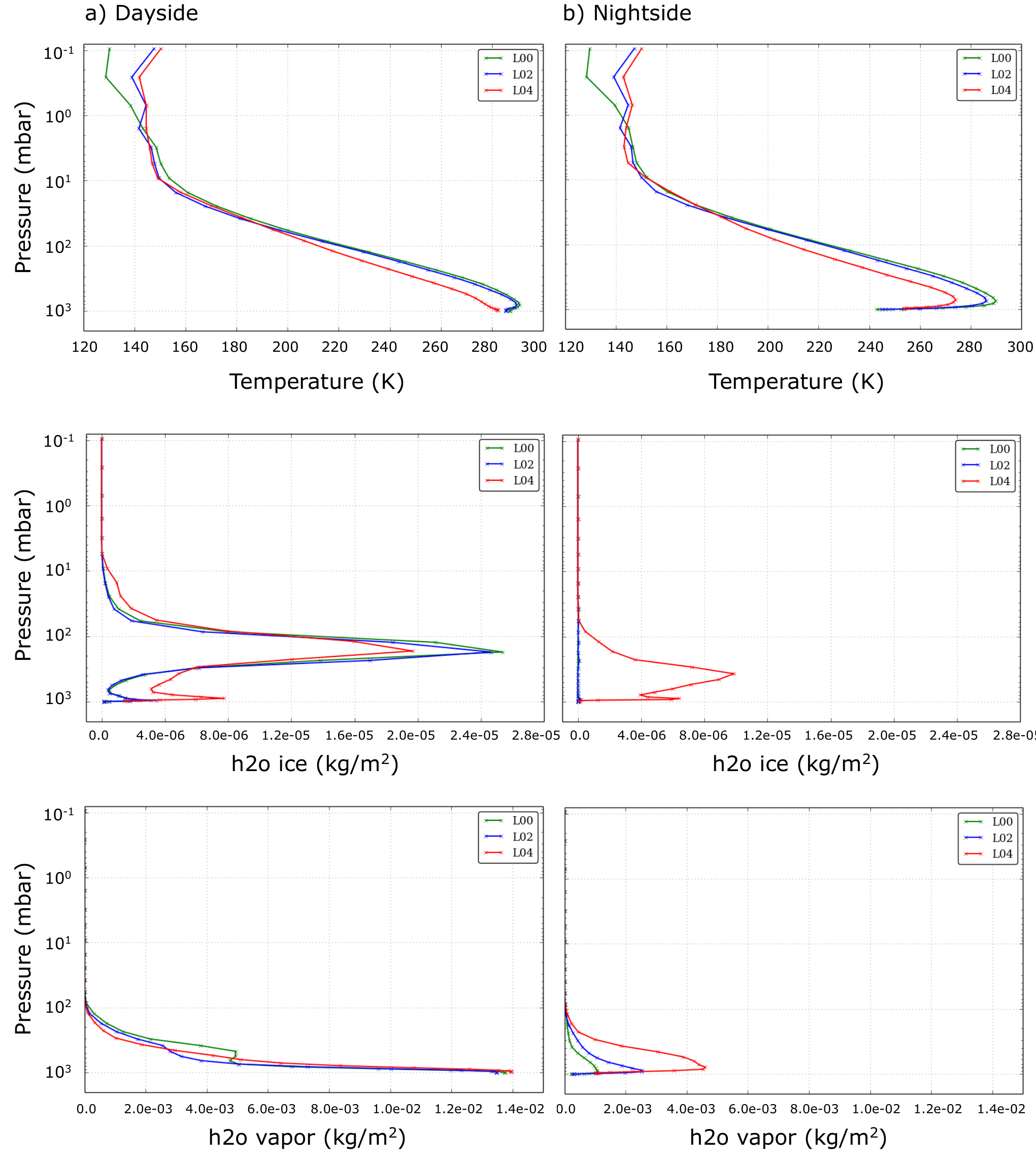}
	\caption{a) Dayside and b) nightside mean profiles of temperature (top panels), water ice (middle panels) and water vapor (bottom panels) in the atmosphere of a planet on a circular orbit around a $1~\Lsun$ star (L00, in green), a $10^{-2}~\Lsun$ star (L02, in blue) and a $10^{-4}~\Lsun$ star (L04, in red).}
	\label{profiles_T_h2oi_h2ov_ecc000_dayside}
	\end{figure*}
	

\section{Circular orbits}
\label{Circular}

In Section \ref{sec_lsun}, we first give our results for a star of luminosity $\Ls = 1~\Lsun$. In Section \ref{lower_lumi}, we compare them with those obtained for a star of luminosity $\Ls = \Lot$ and $\Ls = \Lof$.

\subsection{$\Ls = 1~\Lsun$}\label{sec_lsun}

Figure \ref{L00_ecc000_tsurf_h2o_ice_vap_col_surf} shows the longitude-latitude maps of surface temperature, ice density, water vapor and precipitation on a planet orbiting a Sun-like star after 10,000 days.
The mean surface temperature needs about $\sim 1500$~days to reach its equilibrium, which is about 267~K. 
As the planet is in synchronous rotation, the dayside is much hotter than the nightside (Figure \ref{L00_ecc000_tsurf_h2o_ice_vap_col_surf}, top left). 
The temperatures on the dayside reach 320~K at the substellar point, whereas temperatures on the nightside are around 240~K. 
From an initially free water ocean, an ice cap is formed in a few hundred years (Figure \ref{L00_ecc000_tsurf_h2o_ice_vap_col_surf}, top right). 
About 62\% of the planet is covered with ice, 38\% of the ocean remains free of ice around the substellar point.
We obtain the same kind of features as an eyeball planet \citep[like in][for GJ~581d]{Pierrehumbert2011,Wordsworth2011}.  
Evaporation occurs on a ring around the substellar point (Figure \ref{L00_ecc000_tsurf_h2o_ice_vap_col_surf}, bottom right), and there is a lot of precipitation at the substellar point due to humidity convergence and condensation of moisture along the ascending branch of an Hadley type cell (Figure \ref{L00_ecc000_tsurf_h2o_ice_vap_col_surf}, bottom left). 

The albedo of the planet is about 0.25, which is significantly lower than in \citet{Yang2013}.
This might be due to several differences in our simulations. 
For example, the albedo depends on the size of the cloud-forming ice particules. 
The bigger they are, the lower the albedo. 
In our model, the cloud ice particules have a size varying depending on the water mixing ratio (see \citealt{Leconte2013a} for details). 
In \citet{Yang2013} the size is not indicated but is said to be what is observed on Earth.
Furthermore, \citet{Yang2013} pointed out that the albedo of the planet strongly depends on the oceanic transport, that is not included here. 
Finally, the biggest difference comes from the moist convection parametrization, which is chosen to be very simple with few free parameters in LMDZ \citep{ManabeWetherald1967}. 
This parametrization leads to a lower cloud cover, and thus a lower albedo.

Figure \ref{profiles_T_h2oi_h2ov_ecc000_dayside} shows the dayside and nightside mean profiles of temperature, water ice and water vapor in the atmosphere. 
These mean profiles were obtained by performing a time average over one orbital period\footnote{54 points per orbit were used for the case $\Loo$ (P$_{\rm orb} = 365.5$~day), 122 points for $\Lot$ (P$_{\rm orb} = 22.85$~day) and 98 points for $\Lof$ (P$_{\rm orb} = 1.967$~day). The number of points depends on the output time frequency.}.
The surface dayside temperatures are higher than the surface nightside temperatures. 
Due to this lower surface temperature, we can see a temperature inversion on the nightside, which occurs at a pressure level of about $1$~bar. 
On the dayside, water ice clouds are located around an altitude of $\sim15~\km$ ($\sim 90$~mbar), while on the nightside, there are no water ice clouds in the atmosphere. 
The concentration of water vapor is much higher on the dayside than on the nightside.
For both the dayside and nightside, the water vapor in the atmosphere is essentially within the first 20~km of the atmosphere (pressure $> 0.1$~bar).

On average, the concentration of water vapor in the dayside upper atmosphere (pressure $< 10$~mbar) is about $1\times 10^{-9}~\kg/\m^2$, which is about 100 times more that the water vapor concentration on Earth at the same altitude \citep{Butcher1992}.
It is therefore possible that such planets do experience little water escape as on Earth \citep{Lammer2003, Kulikov2007, Selsis2007}.
Let us note that an extreme case for water escape has been observed for a hot Neptune \citep{Ehrenreich2015}.
However, we expect here a much lower escape rate as the planet is located much farther away.
There are ice clouds above substellar point at an altitude of 15~km, these clouds protect the substellar point. 
This mechanism has been identified by \citet{Yang2013} to allow the inner edge of the insolation HZ for synchronous planets to be closer than for non-synchronous planets.

As a conclusion, over its orbit, a synchronized planet on a circular orbit around a $\Loo$ star has therefore always a part of its ocean ice-free on the dayside (around the substellar point), while its nightside is covered by an ice cap.

\subsection{Decreasing the luminosity}\label{lower_lumi}

Decreasing the stellar luminosity\footnote{We recall that we do not take into account the spectral dependance of a low luminosity star. 
Decreasing the luminosity is done in our work by decreasing the orbital period of the planet, and thus its rotation period (Tables \ref{tab:init1} and \ref{tab:init2}).} changes the global characteristics of the planet's climate, such as the surface temperature map and the surface ice density.
The lower the luminosity the bigger the differences with the previous case.

\subsubsection{$\Ls = \Lot$}

Figure \ref{profiles_T_h2oi_h2ov_ecc000_dayside} shows that there is little difference between the mean temperature profiles at $1~\Lsun$ and $\Lot$.
The temperature at low altitudes (pressure > 2~mbar) is slightly lower but in the upper atmosphere (pressure < 2~mbar) it is bigger (the difference reaches 10 to 20~K).
The water ice clouds follow the same distribution as for a $1~\Lsun$ star, with a high density at an altitude of $\sim15~\km$ ($\sim 90$~mbar) on the dayside and very few clouds on the nightside.  
The water vapor distribution is very similar between the luminosities $\Loo$ and $\Lot$ and the concentration becomes negligible above an altitude of $\sim 20~\km$ (pressure $< 100$~mbar).


	\begin{figure}[htbp!]
	\begin{center}
	\includegraphics[width=9cm]{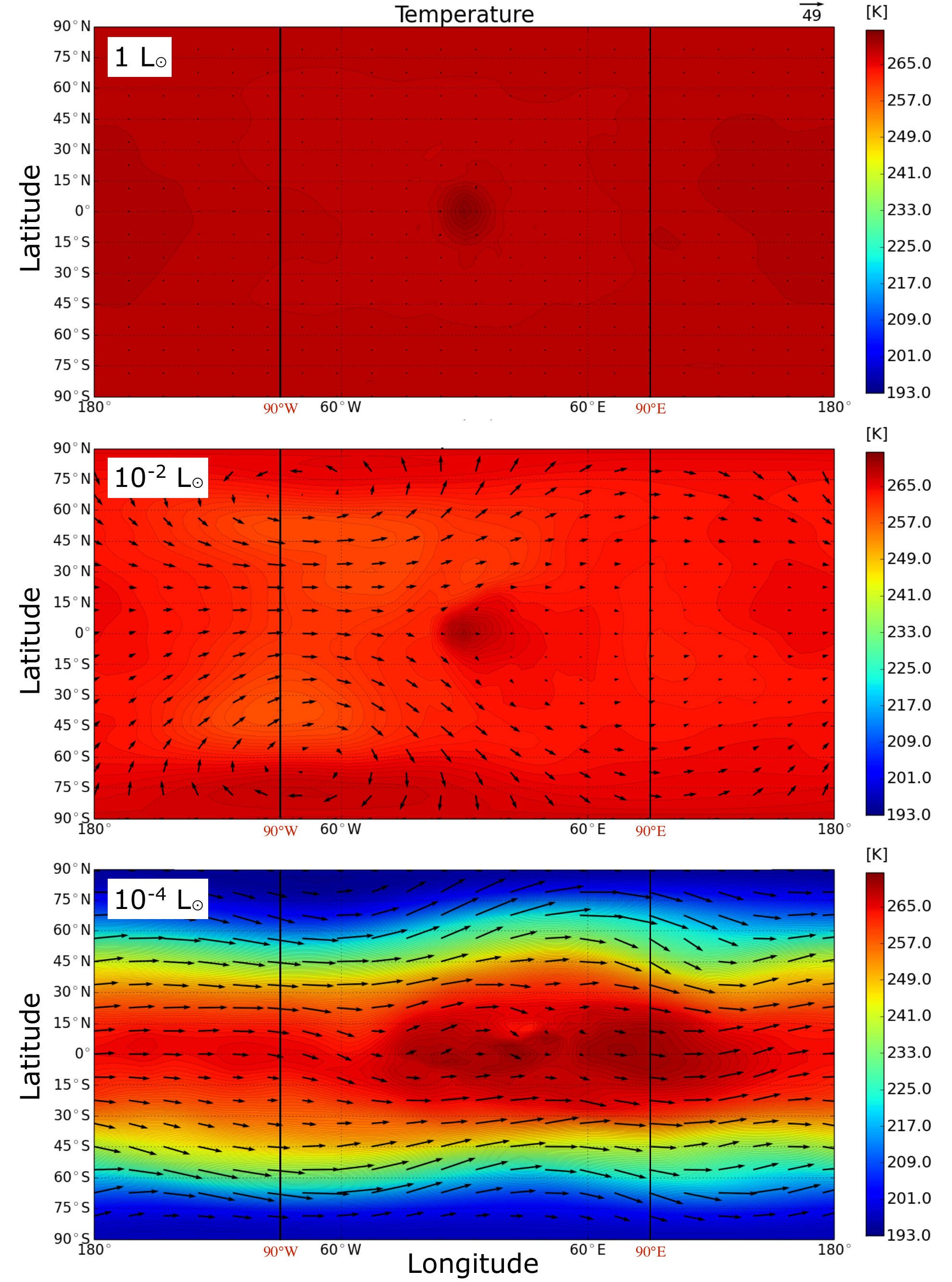}
	\caption{Maps of the atmospheric temperature (color map) and winds (vectors, the units are m/s, legend on the top of the graph) at an altitude of 10~km, for a planet on a circular orbit around a $1~\Lsun$ star (top), a $10^{-2}~\Lsun$ star (middle) and a $10^{-4}~\Lsun$ star (bottom).}
	\label{map_temp_wind_ecc000}
	\end{center}
	\end{figure}
	
However, due to the faster rotation, the temperature distribution is different. 
This is due to the higher Coriolis force that strengthens the mechanisms responsible for the equatorial super-rotation. 
Both the wave-mean flow interaction identified by \citet{ShowmanPolvani2011} and the three-way force balance identified by \citet{Showman2013, Showman2015} affect the atmospheric circulation \citep[see also][for a discussion of the transition from stellar-antistellar circulation to super-rotation in the specific context of terrestrial planets]{Leconte2013a}.
Figure \ref{map_temp_wind_ecc000} shows maps of the atmospheric temperature (color maps) and the wind pattern (with vectors) at an altitude of 10~km (corresponding to a pressure of 305~mbar for $\Loo$, 301~mbar for $\Lot$ and 190~mbar for $\Lof$).
For $1~\Lsun$, the wind is weak and isotropically transports heat away from the substellar point \citep[stellar/anti-stellar circulation, as shown in ][for a slowly rotating Gl-581c]{Leconte2013a}. 
However, for $10^{-2}~\Lsun$, the wind is stronger, especially the longitudinal component, and redistribute more efficiently heat towards the east.
Due to this stronger wind, there is an asymmetry of the surface temperature distribution on the planet, so that the east is hotter than the west and the temperature is more homogeneous along the equator.
The planet orbiting a $10^{-2}~\Lsun$ star is colder at the poles (surface temperature of $\sim 235~$K) than for $1~\Lsun$ star (surface temperature of $\sim 250$~K).
The wind pattern here is marked by the presence of a Rossby wave typical of the Rossby wave transition region \citep[as defined in][for tidally locked planets]{Carone2015}. 
In \citet{Carone2015}, the Rossby wave transition region for a Earth-size planet is said to occur for a rotation period of $25$ day, which is approximately the rotation period here (22.85~day).
In the $\Loo$ case, the rotation period was much longer and no Rossby wave could develop \citep[e.g.,][]{Leconte2013a}.

	\begin{figure}[htbp!]
	\begin{center}
	\includegraphics[width=9cm]{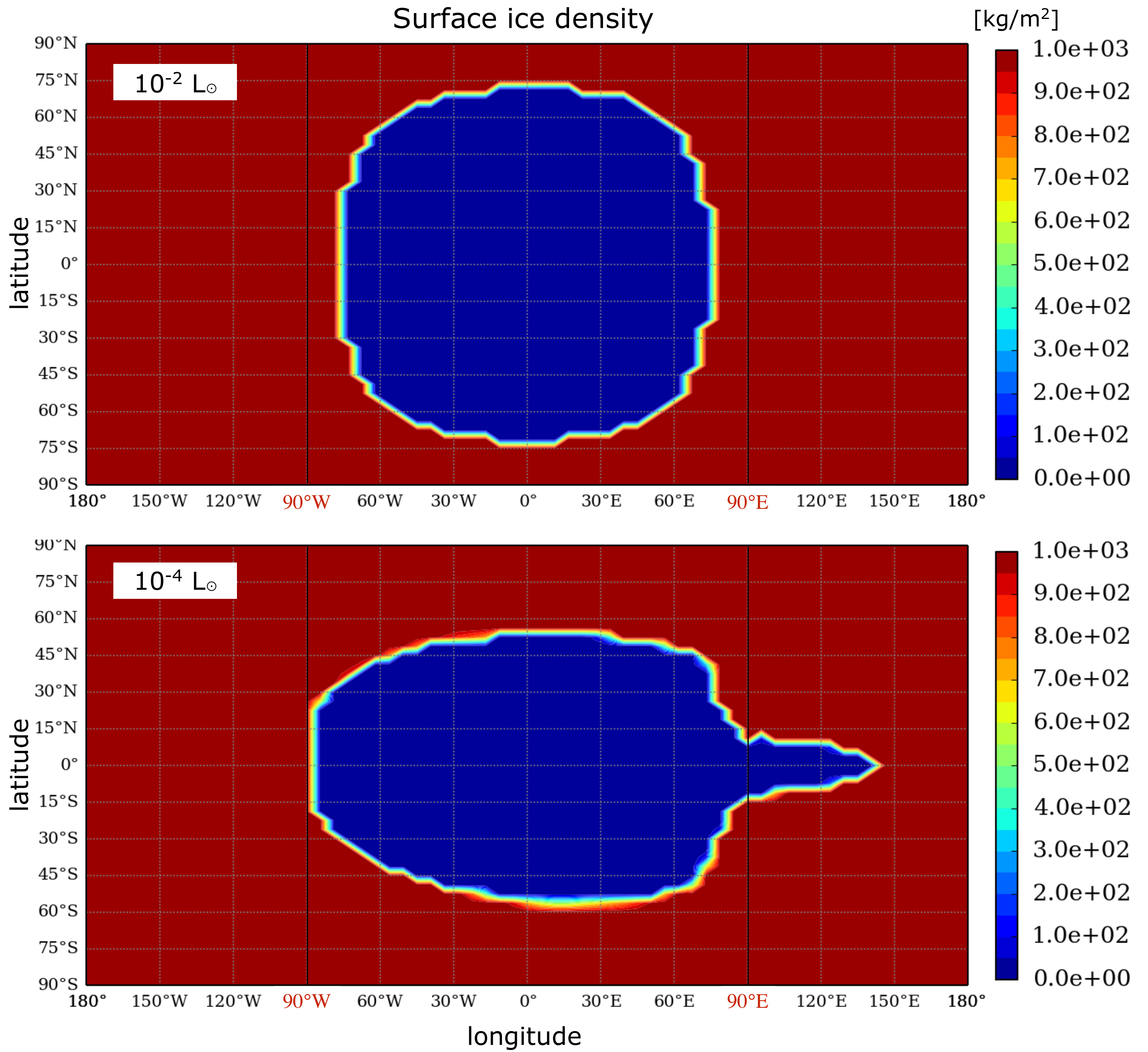}
	\caption{Maps of the surface ice density for a planet on a circular orbit around a $10^{-2}~\Lsun$ star (top) and a $10^{-4}~\Lsun$ star (bottom). The Figure corresponding to the case $\Ls = \Loo$ is in Figure \ref{L00_ecc000_tsurf_h2o_ice_vap_col_surf}.}
	\label{map_h2o_ice_ecc000_bis}
	\end{center}
	\end{figure}

Moreover, for $\Lot$, the averaged ice density is similar to the one of the planet orbiting a $1~\Lsun$ star.
However, due to the asymmetric surface temperature map due to the Coriolis force, the shape of the ice-free region is slightly different.
Figure \ref{map_h2o_ice_ecc000_bis} shows the shape of the surface ice density for $\Ls = \Lot$ (top panel).
Although the percentage of ice-free region remains the same, the ocean for $\Lot$ reaches slightly lower latitudes and is more extended longitudinally than for $\Loo$.

Despite these small changes with respect to the case $\Loo$, a planet on a circular orbit around a $\Lot$ star is therefore equally capable to sustain surface liquid water.

\subsubsection{$\Ls = 10^{-4}~\Lsun$}

For an even less luminous host body, the differences that were appearing for the case $\Ls = 10^{-2}~\Lsun$ are accentuated. 
The temperature maps and surface ice density maps show a more longitudinal extension than before.
Due to the even higher rotation rate of the planet (rotation period of 2~day, see Table \ref{tab:init2}), the winds are redistributing heat longitudinally from the substellar point, heating up efficiently the nightside of the planet. 
Figure \ref{map_temp_wind_ecc000} shows that the winds are much stronger for the case $\Lof$ ($\sim 50~$m/s vs $\sim 25~$m/s for $\Lot$ and $< 10~$m/s for $\Lof$) and redistribute very efficiently the heat along the equator.
In this case, we notice the presence of longitudinal wind jets at latitudes of $45\deg$S and $45\deg$N and at a pressure level of $\sim 30$~mbar.
The wind velocity in the jets can reach almost 120~km/h.
As the rotation period is of $\sim2$~day, the criteria for the Rossby wave to be triggered is met and super-rotation takes place \citep{ShowmanPolvani2011, Leconte2013a, Carone2015}.
As for $\Lot$, the colder regions are the poles, but their temperature is here even lower (a surface temperature about 20~K lower than for $\Lot$).

Figure \ref{profiles_T_h2oi_h2ov_ecc000_dayside} shows that this super-rotation also causes a longitudinal uniformisation of the temperature, water ice clouds and water vapor distribution.
Indeed, the nightside temperature of the lowest layer of the atmosphere for $\Lof$ is not as low as for $\Loo$ and $\Lot$, and there are much more ice clouds and water vapor on the nightside for $\Lof$ than for $\Loo$ and $\Lot$.

Let us note that the evolution of the surface ice density is less quick than in the previous two cases, for which in less than a few decades the surface ice density reaches its equilibrium.
Here a few thousands days are needed to reach equilibrium. 
One of the main difference it implies is an initially lower surface ice density for planets around $\Ls = \Lof$ stars than for more luminous objects. 
First, the ocean free region survives and forms a belt around the equator with the belt buckle at substellar point. 
As the eastward wind is losing heat, the eastern regions are hotter than the western regions.
After about 700~days of evolution, the ice forms a bridge at 120$\deg$ west, closing the equatorial ocean.
Figure \ref{map_h2o_ice_ecc000_bis} shows that, when the equilibrium is reached, the shape of the ocean for $\Ls = \Lof$ is very different from the ones of the other two luminosities.
When the equilibrium is reached, we find that this planet has an ice-free ocean of a similar size as the previous cases: about 40\% of the planet is ice-free.

Figure \ref{comp_mean_tsurf_e01_zoom} shows the mean surface evolution and mean ice density evolution for planets around the three different host stars.
For the circular orbits, we can see that the mean surface temperature is higher for $\Ls = \Lof$ than for $\Ls = \Loo$ and $\Ls = \Lot$.
This is due to the strong winds in the atmosphere of the planet, which chase the clouds that are forming above the substellar point.
Without the cloud protection, the surface temperature increases. 
For a tidally locked planet orbiting very low luminosity objects, the longitudinal winds are strong and the stabilizing cloud feedback identified by \citet{Yang2013} is less efficient. 
Therefore, one might expect that the HZ of tidally locked planets around very low luminosity stars is not as extended as in \citet{Yang2013}.
However, simulations for higher incoming flux should be performed to verify this claim, in particular to see how the albedo varies with increasing incoming flux.


\section{Eccentric orbits}
\label{Eccentric}

As described in Section \ref{initial_conditions_planets}, we made sure that all planets receive on average the same flux as Earth.
Increasing the eccentricity changes significantly the insolation the planet receives over one orbit and it leads to changes in general characteristics of the climate.
The insolation varies over the orbit and the substellar point moves along the equator due to the optical libration.
Temperature, atmospheric water ice, atmospheric water vapor and surface ice are influenced by this forcing, as shown in the following.

\begin{figure}[htbp!]
	\begin{center}
	\includegraphics[width=9cm]{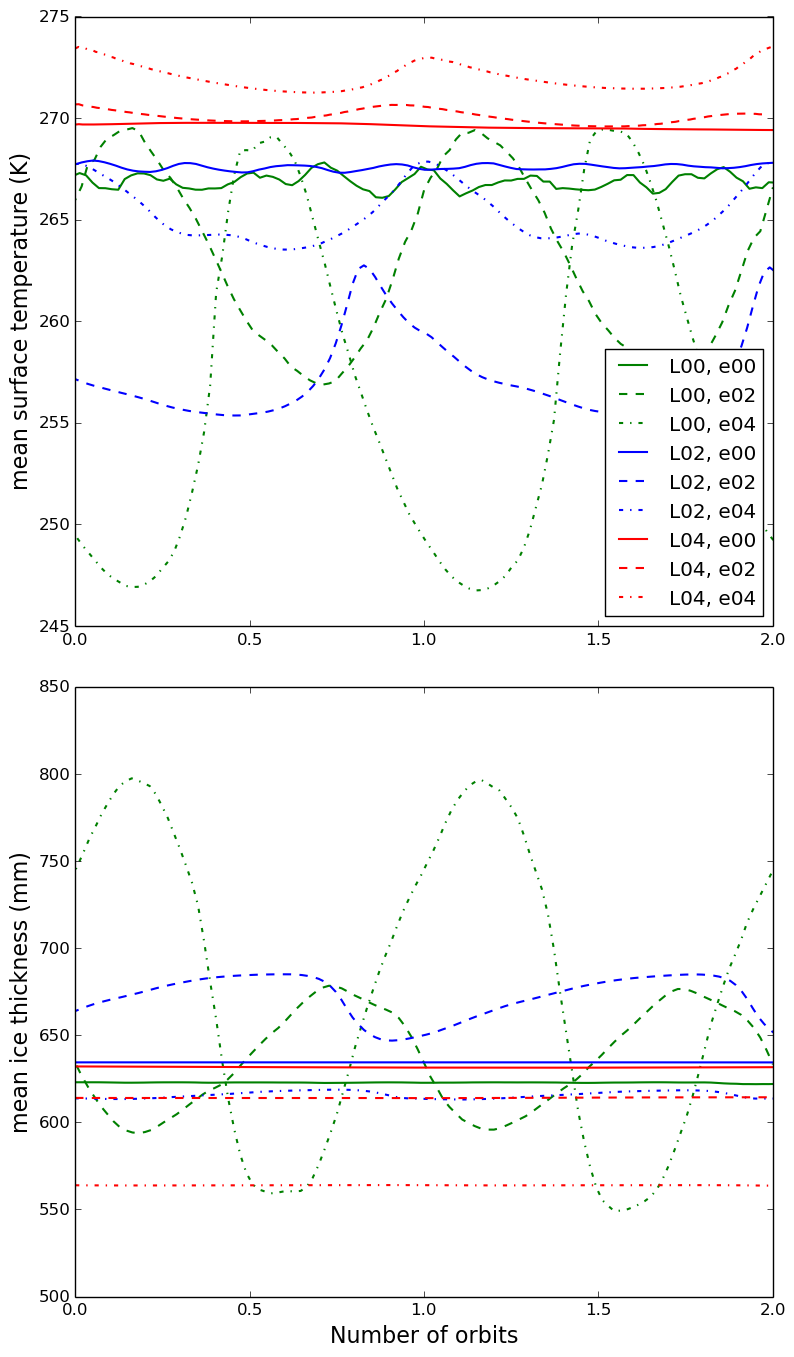}
	\caption{Evolution over two orbits of the mean surface temperature and the mean ice thickness of a planet orbiting a $\Loo$ star (L00, green), a $\Lot$ star (L02, blue) and a $\Lof$ (L04, red) star on an orbit of eccentricity 0.0 (e00), 0.2 (e02) and 0.4 (e04). Note that the orbital periods are different for each case (see Tables \ref{tab:init1} and \ref{tab:init2}).}
	\label{comp_mean_tsurf_e01_zoom}
	\end{center}
	\end{figure}

Figure \ref{comp_mean_tsurf_e01_zoom} shows an example of the evolution, during two orbits, of the mean surface temperature and mean ice thickness for planets on a circular orbit and planets with an eccentricity of 0.2 and 0.4, orbiting the three different kinds of object.
The mean surface temperature oscillates with the change of insolation over the orbit.
As the orbital period changes with the luminosity and eccentricity (see Tables \ref{tab:init1} and \ref{tab:init2}), it oscillates with different frequency and amplitude for the different cases.
This oscillation of insolation and thus temperature has an effect on the amount of water vapor and surface ice (as seen in Figure \ref{comp_mean_tsurf_e01_zoom}).
We discuss in the following sections how it has an impact on the climate of the planets considered here and the presence of liquid water at their surface.

\subsection{Luminosity of $\Ls = 1~\Lsun$}

First, we consider a luminosity of the star similar to the luminosity of the Sun.
Firstly, for small eccentricities, our simulations show surface temperature oscillations, ice density and water vapor oscillations. 
However, they remain small enough for the planet to keep surface liquid water.
For example, a planet on an orbit of eccentricity $0.2$ experiences temperature oscillations on the dayside of about 30~K in 371~days, while the mean temperature oscillations are of about 12~K (see Figure \ref{comp_mean_tsurf_e01_zoom}).
The surface ice density responds accordingly with the eccentricity-driven seasonal melting and freezing. 
On average, the surface ice density varies between $\sim 58$\% after the passage at periastron and $\sim 66$\% after the passage at apoastron.
The region around the substellar point is always ice-free but the center of the ice-free region shifts on the surface of the planet by about 10$\deg$ during the orbit.

	\begin{figure*}[htbp!]
	\centering
	\includegraphics[width=14cm]{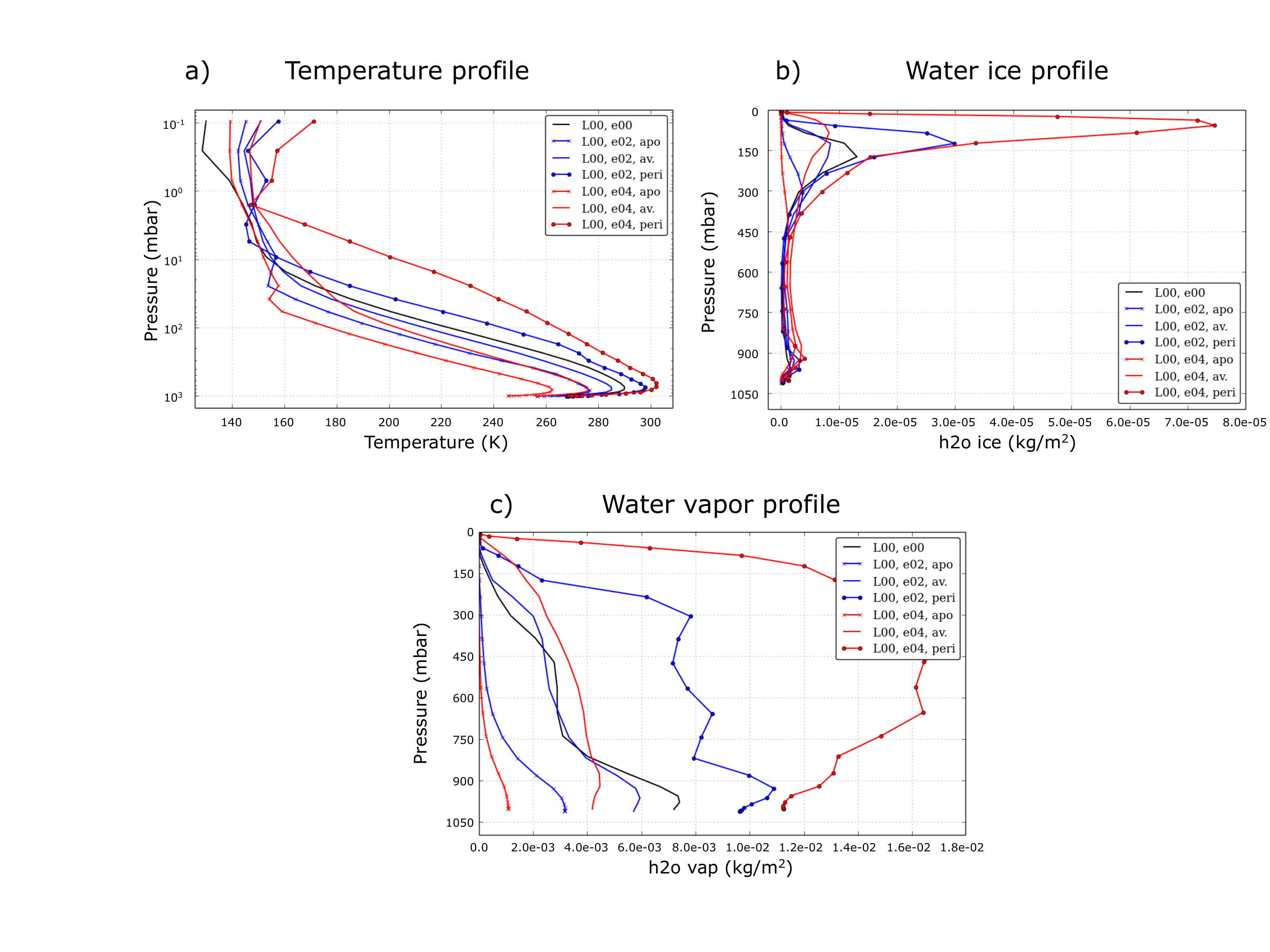}
	\caption{Mean profiles of a) temperature (logarithmic scale), b) water ice and c) water vapor (both in linear scale) in the atmosphere of a planet orbiting a $1~\Lsun$ star, on a circular orbit (black), an orbit of eccentricity 0.2 (blue) and 0.4 (red). For the eccentric cases, the temperature profiles at periastron (peri) and apoastron (apo) are represented as well as the time average over one orbit (av.).}
	\label{profiles_T_h2oi_h2ov_L00}
	\end{figure*}

Secondly, when we increase the eccentricity to 0.4, the amplitude of the variations is higher.
For example, Figure \ref{comp_mean_tsurf_e01_zoom} shows that the mean temperature variations increase from $\sim 12$~K for an eccentricity of $e=0.2$ to $\sim23$~K for an eccentricity of $e=0.4$. 
Figure \ref{comp_mean_tsurf_e01_zoom} also shows that the amplitude of the mean ice density variations of the planet also increases with the eccentricity. 
The planet's surface ice density after periastron is lower than previously ($\sim 55$\% vs. $\sim 58$\%), but is much bigger after apoastron ($\sim 80$\% vs. $\sim 66$\%).
The planet is never completely frozen during its evolution since an ice-free region always survives even after the passage at apoastron.
Figure \ref{profiles_T_h2oi_h2ov_L00} shows the mean profiles of temperature, water ice and water vapor in the atmosphere for different eccentricities. 
For non zero eccentricities, the different quantities are plotted around apoastron and around periastron\footnote{Actually, it is a few days after the passage at periastron or apoastron, as the atmosphere responds with a lag. We select the extreme values of the atmospheric water ice and vapor.}.
When the eccentricity increases, the apoastron and periastron profiles depart more from the circular mean profile. 
At apoastron, the temperature profile is colder than for $e=0$. 
There are less ice water clouds in the atmosphere and they are located closer to the surface. 
There is also less water vapor in the atmosphere.
On the contrary, at periastron, the temperature profile is hotter, there are much more clouds located higher in the atmosphere and there is also more water vapor in the atmosphere.
For an eccentricity of 0.4, the water vapor concentration in the upper atmosphere can reach a few $10^{-7}~\kg/\m^2$, which is about 7000 times more that the water vapor concentration on Earth at the same altitude \citep{Butcher1992}.
Due to the passage at periastron where the planet can receive up to 2.5 times the insolation flux of the Earth, the water vapor peaks in the high atmosphere (pressure < 1~mbar) for about 200~days above $10^{-8}~\kg/\m^2$, and for about 30~days above $10^{-7}~\kg/\m^2$.
There is more water in the upper atmosphere for a planet on an orbit of eccentricity 0.4 than there is in the upper atmosphere of a planet on a circular orbit.

We thus expect atmospheric loss to be more important for a planet on an eccentric orbit than a planet on a circular orbit. 
This process happens faster at periastron where the star-planet distance is shorter, which also coincides with the moment when the concentration of water vapor in the atmosphere is higher. 
We thus expect a larger atmospheric escape rate at periastron.
However, as atmospheric escape happens on very long timescales, it might be sensitive only to averaged value of the water vapor concentration, the difference of water vapor concentration at periastron or apoastron could not matter.

\begin{figure}[htbp!]
	\begin{center}
	\includegraphics[width=9cm]{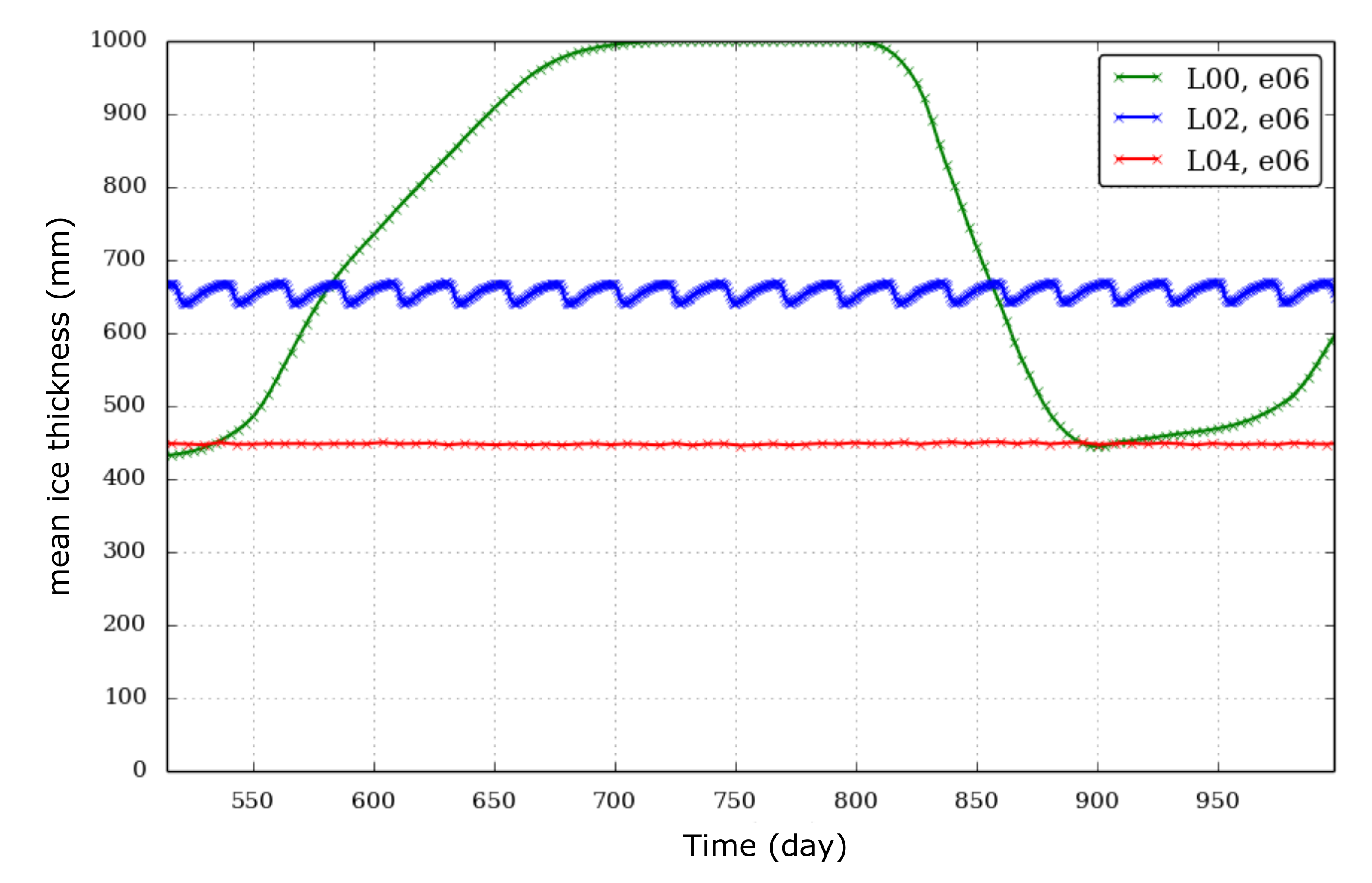}
	\caption{Evolution of the mean ice thickness (here a thickness of $1000$~mm means the whole planet is covered by the maximum depth of ice allowed in our model: $\hice =1$~m) of a planet orbiting a $\Loo$ star (green), a $\Lot$ star (blue) and a $\Lof$ star (red) on an orbit of eccentricity 0.6.}
	\label{mean_h2o_ice_surf_ecc060}
	\end{center}
	\end{figure}

Thirdly, for eccentricities higher than 0.6, the planet is completely frozen around apoastron (corresponding to a mean ice thickness of 1~m, our value of $\hice$.). 
Figure \ref{mean_h2o_ice_surf_ecc060} shows the evolution of the mean ice thickness for the three different host bodies.
For an eccentricity of 0.6, the orbital period is 432 days and the planet spends $\sim100$ days in a completely frozen state, which corresponds to about $20$\% of its orbital period.
When the planet gets closer to the periastron, the ice starts melting at a longitude of 60$\deg$ West and liquid water is again available on the dayside around periastron.
However the water vapor concentration becomes very high and the surface temperature also increases to more than 300~K.

\medskip

As a conclusion, for planets with an eccentricity less than 0.6, the planet is always able to sustain surface liquid water. 
The ice-free region changes and shifts as the planet revolves around the star but never disappears.
For planets with an eccentricity above 0.6, the planet cannot sustain surface liquid water during the whole orbital period.
Let us note that for very high eccentricities and around periastron, the temperatures become higher than what the model allows us ($> 400~K$).
The mean flux approximation is therefore less valid for planets orbiting Sun-like stars on very high eccentricity orbits. 
However, departing from our definition of habitability based on sea-ice cover, we could speculate about how a potential life formed on such a planet would survive this succession of frozen winters around apoastron and hot summers around periastron.

\subsection{Decreasing the luminosity}

Decreasing the luminosity has the effect of decreasing the eccentricity-driven insolation oscillation period. 
Indeed, the insolation varies on a shorter timescale and this affects the climate response.
The lower the luminosity, the less time the climate has to respond to the forcing.

\subsubsection{Luminosity of $\Ls = \Lot$}

For a planet orbiting a $\Lot$ star with an eccentricity of 0.2, the dayside experiences temperature fluctuations of $\sim 40~\K$ in about 23~days.
It is of the same order of magnitude as for a $\Loo$ star, but the fluctuation happens on a much shorter timescale.
Recall that decreasing the luminosity was done by decreasing the orbital period (the planets have to be closer to receive a flux of $1366$~W/m$^2$), and as the planets are synchronized, the rotation period decreases as well (see Table \ref{tab:init2}). 
We did not take into account the spectral dependance of the low luminosity stars.


However, the oscillations of the average quantities have a much smaller amplitude than for $\Loo$.
Indeed, Figure \ref{comp_mean_tsurf_e01_zoom} shows that the amplitudes of oscillations of mean surface temperature and mean ice thickness are significantly damped when decreasing the luminosity from $\Loo$ to $\Lot$.
Due to the shorter orbital period, the frequency of the forcing is higher than for the case $\Loo$ and the oscillations in the mean temperature, mean ice thickness, and mean water vapor concentration have a lower amplitude.
The climate has indeed less time to react to the insolation forcing.

\begin{figure}[htbp!]
	\begin{center}
	\includegraphics[width=9cm]{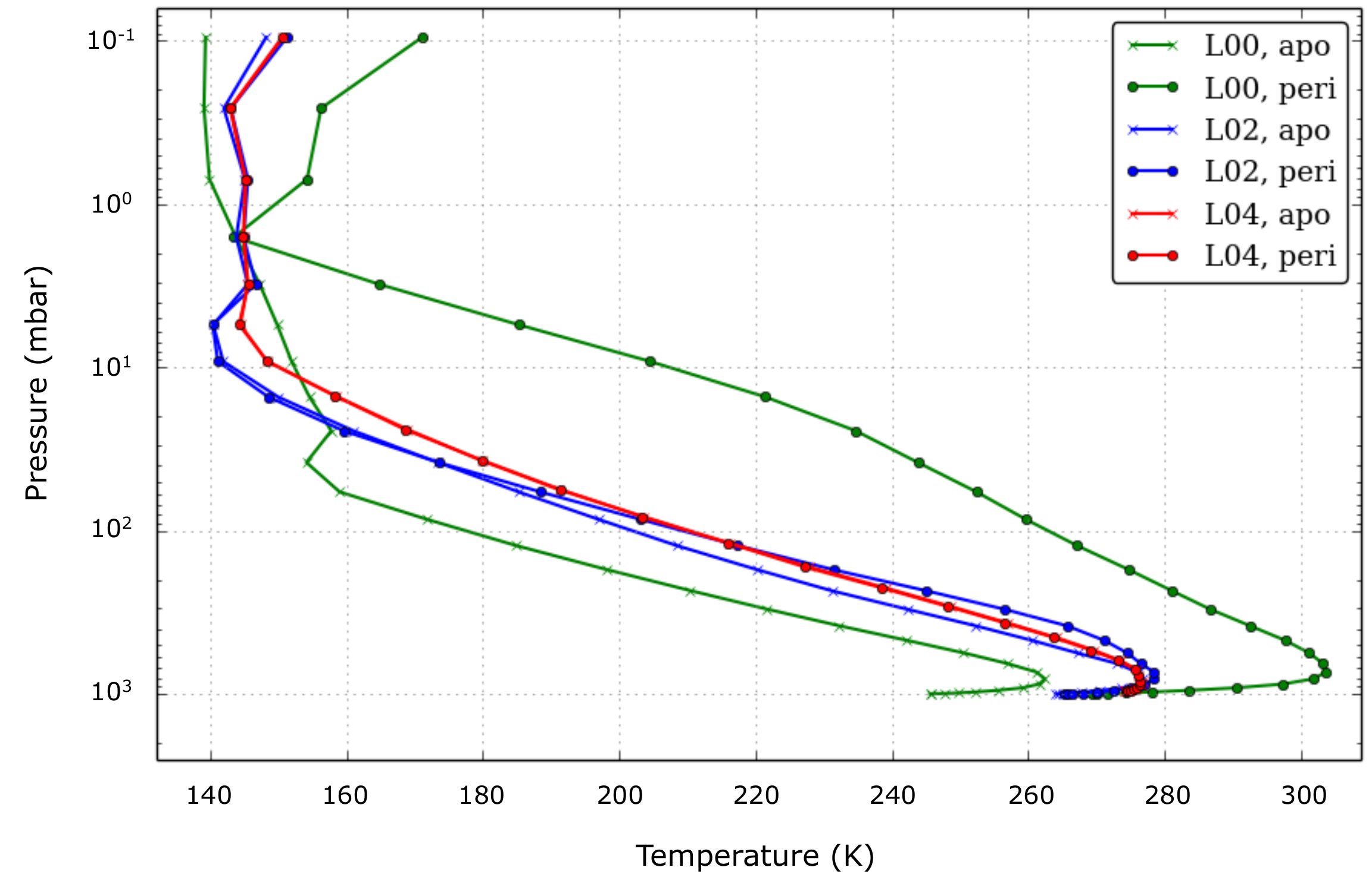}
	\caption{Mean atmospheric temperature profile for planets with an eccentricity of 0.4, orbiting a star of $\Loo$ (green), $\Lot$ (blue) and $\Lof$ (red), for periastron and apoastron. Note that the two red curves are superimposed.}
	\label{temp_intercomp_ecc040}
	\end{center}
	\end{figure}

Figure \ref{temp_intercomp_ecc040} shows the mean atmospheric temperature profile for a planet at periastron and apoastron of an orbit of eccentricity 0.4 around the different kinds of object. 
The difference between apoastron and periastron is important for the case $\Loo$ with a difference of about 40~K at pressure level of 800~mbar.
However, for $\Lot$ the difference is negligible: a few kelvins from a pressure level of 800~mbar to 50~mbar and in the upper atmosphere (pressure < 1~mbar). 
This shows the slowness of the climate to respond to the higher-frequency forcing.

	\begin{figure}[htbp!]
	\begin{center}
	\includegraphics[width=9cm]{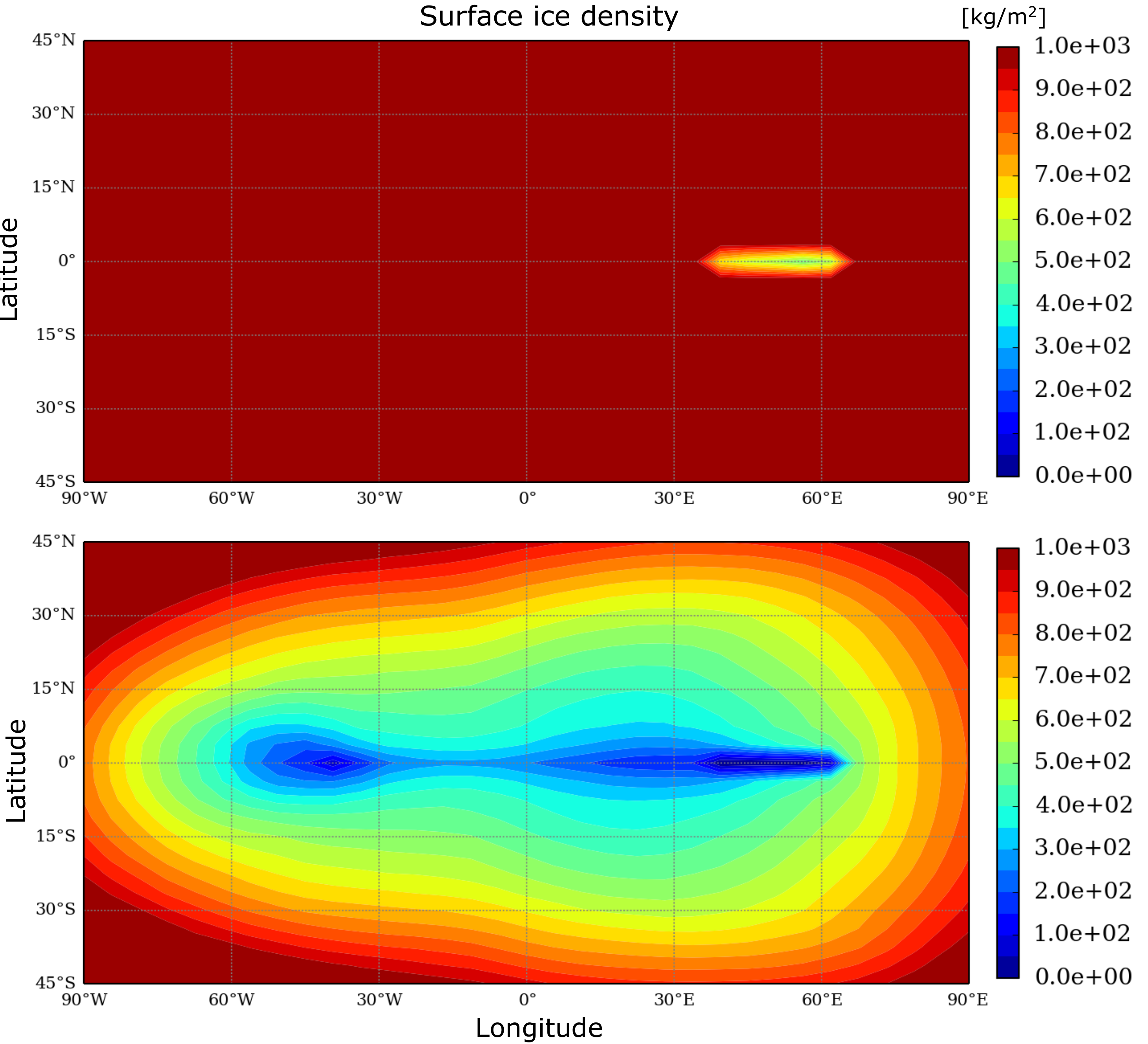}
	\caption{Surface ice density for a planet of eccentricity 0.8 orbiting a $\Lot$ star. Top graph: when the surface ice density is maximum (just before the periastron passage, i.e. after the long eccentricity-induced winter), bottom graph: when the surface ice density is minimum (just after the periastron passage).}
	\label{map_h2o_ice_L02_ecc080_apo_peri}
	\end{center}
	\end{figure}
	
Decreasing the luminosity has the effect of pushing the limit of liquid water presence towards higher eccentricities.
For example, Figure \ref{mean_h2o_ice_surf_ecc060} shows that a planet with an eccentricity of 0.6 orbiting a $\Lot$ star can always sustain surface liquid water whereas it cannot do so for the whole orbital period if it orbits a $\Loo$ star.

For an eccentricity of 0.8, the planet passes by a complete frozen state around apoastron and partially melts around periastron so that for about ten days ($\sim 1/3$ of the orbit), there is a small ice-free oblong region.
This state is reached in $\sim7500$~days.
Figure \ref{map_h2o_ice_L02_ecc080_apo_peri} shows the surface ice density on the planet around periastron. 
Just after periastron, less than a percent of the planet's surface is ice-free at the equator around a longitude of $50\deg$E (with an extent of about ten degrees in longitude and less than a degree in latitude, see bottom panel of Figure \ref{map_h2o_ice_L02_ecc080_apo_peri}).
Just before periastron, i.e. after the long eccentricity-induced winter, the region that was ice-free around periastron freezes but the ice layer depth does not reach the maximum value of 1~m (see top panel of Figure \ref{map_h2o_ice_L02_ecc080_apo_peri}).
However, for an eccentricity of 0.9, the planet rapidly becomes completely frozen. 
This state is reached at the first passage at apoastron, where the planet freezes completely with 1~m of ice covering the whole planet and melts partially around periastron. 
However, even around periastron there is always a thin layer of ice covering the whole planet.


\medskip

As a result, we find that planets orbiting a $\Lot$ star can always sustain surface liquid water on the dayside for higher eccentricities than for a $\Loo$ star (up to $0.8$, instead of $0.6$).
For an eccentricity of 0.8 and higher, the planet remains completely frozen.
All in all, the climate simulations are more in agreement with the mean flux approximation for planets orbiting $\Lot$ stars. 
This is due to the averaging of the climate caused by the quicker rotation. 
However, departing from our definition of habitability based on sea-ice cover, we could speculate about how a potential life form on such a planet would survive this rapid succession of frozen winters and hot summers (several tens of Kelvins in only $\sim 20$~days).

\subsubsection{Luminosity of $\Ls = \Lof$}

For an even less luminous host body, the amplitude of the oscillations in mean temperature and mean ice thickness are damped with respect to the other cases. 
As the duration of the year is very short (2 to 4~days, as specified in Table \ref{tab:init2}), there is therefore a more effective averaging of the mean surface temperature and ice thickness than for longer orbital period planets.
Figure \ref{temp_intercomp_ecc040} shows the mean temperature profile for planets at periastron and apoastron of an orbit of eccentricity 0.4. 
For the case $\Lof$, there is no difference between periastron and apoastron. 
The shape of the surface ice density also changes when the luminosity decreases. 
For $\Lot$, the ice-free region has a oblong shape, for $\Lof$ it has a more peanut shape due to the passage at periastron. 
The mean ice thickness for $\Ls = \Lof$ is similar to the two previous cases but does not significantly vary over time (see Figure \ref{mean_h2o_ice_surf_ecc060}). 
  
\medskip

We find that planets orbiting a $\Lof$ star remain always capable to sustain surface liquid water on the dayside throughout the orbit up to very high eccentricities (0.9, instead of 0.8 for $\Lot$ and 0.6 for $\Loo$), due to the efficient averaging of the climate brought up by the rapid rotation. 
The mean flux approximation is therefore valid for small luminosities.  
However, for an eccentricity of 0.8, the dayside temperature varies over $\sim 50$~K, between 300~K and 350~K in less than 4 days.
This situation is very extreme for potential life so that the question remains of how life could appear in such an unstable environment.

\section{Changing the thermal inertia of the oceans $I_{oc}$ and the ice thickness $\hice$}
\label{therm_inert_hice}

\begin{figure}[htbp!]
	\begin{center}
	\includegraphics[width=9cm]{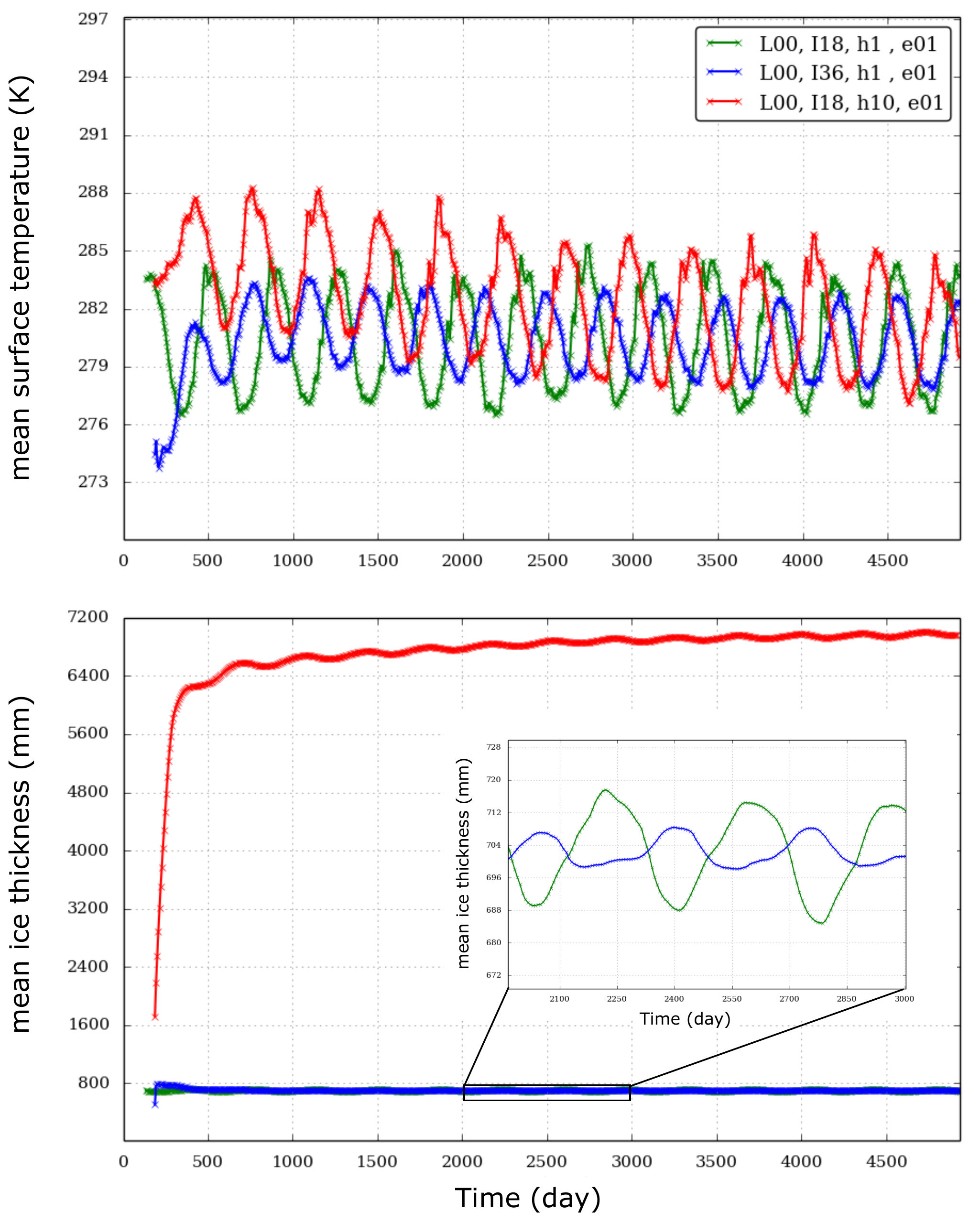}
	\caption{Mean surface temperature and mean ice thickness of a planet orbiting a $\Loo$ star with an eccentricity of $0.1$. Our standard scenario is in green: $I_{oc} = 18000~\inertia$, $\hice = 1~\m$. $I_{oc} = 36000~\inertia$ and $\hice = 1~\m$ is represented in blue and $I_{oc} = 18000~\inertia$, $\hice = 10~\m$ in red.}
	\label{comp_Ih_tsurf_ice_L00_e01_article}
	\end{center}
	\end{figure}

Changing the thermal inertia of the oceans or the ice thickness modifies a few properties of the planets climate but the consequences for surface liquid water presence remain basically the same. 
The time needed to reach the equilibrium does not change significantly when increasing the thermal inertia of the ocean from $18000~\inertia$ to $36000~\inertia$ but, as Figure \ref{comp_Ih_tsurf_ice_L00_e01_article} shows, it increases when increasing the ice thickness from 1~m to 10~m.
Indeed, as the model is allowed to create thicker ice layers, more time is needed to reach the equilibrium ice thickness. 

\medskip

For planets orbiting a $\Loo$ star on a circular orbit, changing the thermal inertia does not influence the overall evolution. 
The mean surface temperature evolves similarly for both thermal inertia and the equilibrium value is slightly higher for $I_{oc} = 18000~\inertia$.
The ice is more present for $I_{oc} = 18000~\inertia$ than for $I_{oc} = 36000~\inertia$. 
When increasing the eccentricity, the main difference is that the oscillation amplitude in mean surface temperature and mean ice thickness is lower for $I_{oc} = 36000~\inertia$.
For an eccentricity of 0.4, the mean surface temperature fluctuations amplitude is divided by two when multiplying the thermal inertia by two.   

With a higher thermal inertia, the oceans can damp the climate fluctuations more efficiently.
Although the temperature fluctuations on the dayside of the planet are still important, the extent of the ocean varies less with $I_{oc} = 36000~\inertia$ than with $I_{oc} = 18000~\inertia$, which makes it an environment more stable for potential life. 
However, this higher thermal inertia of the oceans does not prevent the planet to be completely frozen for very eccentric orbits. 
For example, for $e=0.8$, the planet freezes completely at apoastron. 

For planets orbiting a $\Lot$ star on a circular orbit, the results are similar to the case $\Loo$. 
However, the damping of the oscillations amplitude is not as pronounced.
Due to the faster rotation of the planet and the more efficient heat redistribution, the planet climate's is more averaged and is less sensitive to the difference in thermal inertia.   

Increasing the thermal inertia of the ocean allows for the climate to have less extreme variations. 
Following our definition of habitability, changing the thermal inertia of the oceans does not influence the results. 
While the planet is able to sustain surface liquid water on the dayside for $I_{oc} = 18000~\inertia$, it still can do so for $I_{oc} = 36000~\inertia$. 
While the planet is only able to temporally sustain surface liquid water on the dayside around periastron for $I_{oc} = 18000~\inertia$, it is also the case $I_{oc} = 36000~\inertia$.

\medskip

For planets orbiting a $\Loo$ star on a circular orbit, changing the maximum ice thickness, $\hice$, allowed in the model does change neither the equilibrium temperature of the planet (see Figure \ref{comp_Ih_tsurf_ice_L00_e01_article}), nor the amount of water vapor in the atmosphere, nor the geographical repartition ocean-ice.
However, more time is needed to reach the equilibrium because a thicker ice-layer has to be created.

For eccentric orbits, the eccentricity-driven oscillations of mean surface temperature and mean water vapor content are not damped with respect to the case $\hice = 1~\m$.
However, the shape of the ocean free zone remains very similar throughout the eccentric orbit. 
The ice layer on the border of the ice-free zone located in the western and eastern regions, which receives a strong insolation when the planet passes by periastron, has time to only partially melt down.
The ice-free zone is therefore smaller than for $\hice = 1~\m$.

\section{Observables}\label{Observables}

The orbital phase curves of eccentric exoplanets have already been observed-  for instance for the close-in giant planets HAT-P-2B by \citet{Lewis2013},  HD~80606 by \citet{Laughlin2009} or WASP-14b by \citet{Wong2015} - and modeled \citep[e.g.][]{IroDeming2010, Lewis2010, KaneGelino2011, CowanAgol2011, Selsis2013, Kataria2013, Lewis2014}. 
We show here that the variability caused by the eccentricity can be observed by orbital photometry in the visible as well as in the thermal infrared.

The model computes the top-of-the-atmopshere (TOA) outgoing flux in each spectral band and for each of the 64x48 columns. 
The flux received from the planet by a distant observer at a distance $d$ in the spectral band $\Delta \lambda$  is given by: 
$$ \phi_{\Delta \lambda}(d) = \sum_{lon,lat} \frac{F_{\Delta \lambda,lon,lat}}{\pi} \times \frac{S_{lat} ~\mu_{lon,lat}}{d^2},$$
where the first term in the sum is the top-of-the-atmopshere scattered or emitted intensity, which is assumed to be isotropic, and the second term is the solid angle subtended by an individual cell. 
$S_{lat}$ is the area of the cell and $\mu_{lon,lat}$ is cosinus of the angle between the normal to the cell and the direction toward the observer. 
If this angle is negative then the cell is not on the observable hemisphere of the planet and we set $\mu_{lon,lat}=0$. 
In practice we use a suite of tools developed by \citet{Selsis2011} that can visualize maps of the emitted/scattered fluxes and produce time- or phase-dependent disk-integrated spectra at the spectral resolution of the GCM. 

	\begin{figure}[htbp!]
	\begin{center}
	\includegraphics[width=\linewidth]{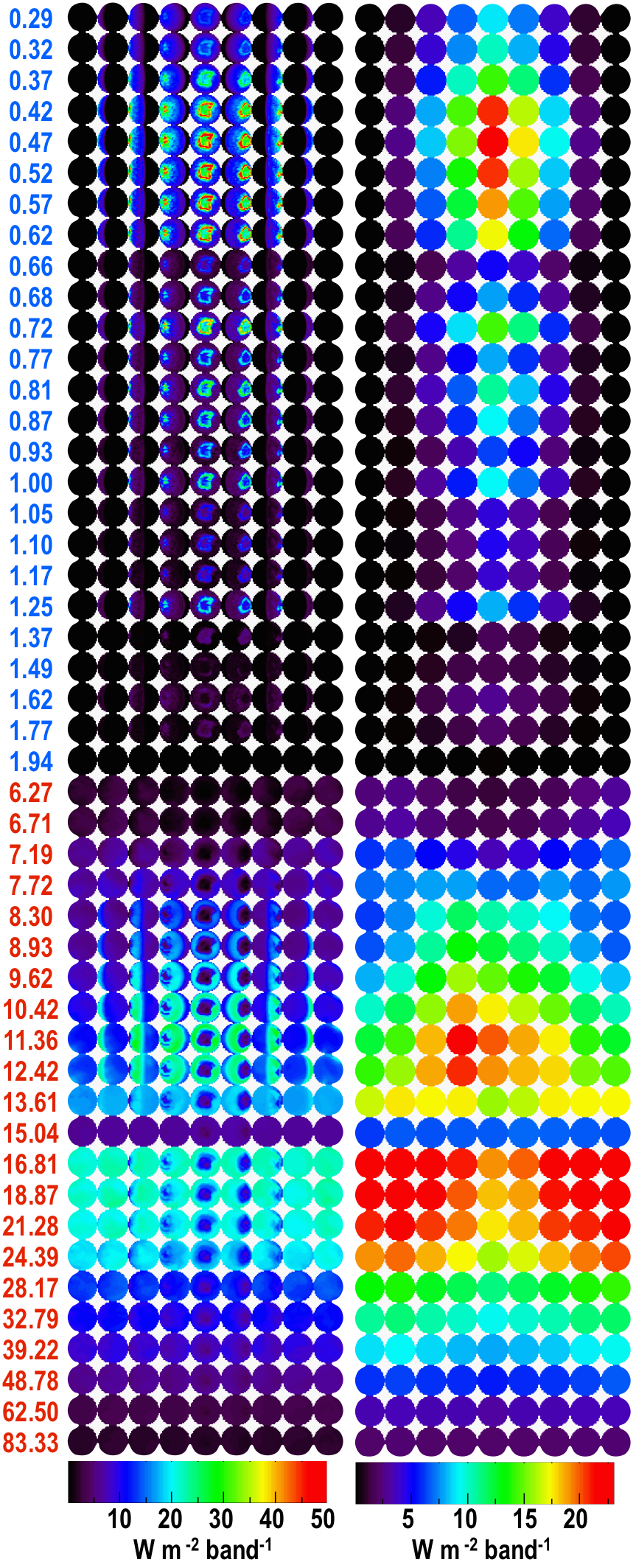}
	\caption{Maps of TOA fluxes as seen by a distant observer for one given orbit and one observation geometry ($\Loo$, circular case). Each line represents one full orbit observed in one band (with the superior conjunction at the center). Each column represents a spectrum of the planet at the GCM resolution at a given orbital phase. The central wavelength of the bands is given in $\mu$m on the left. Note the gap between short (scattered light, in blue) and long (thermal emission, in red) wavelengths. On the right panel, the averaged spatially-unresolved flux is given with a different color scale (due to a smaller range of values). }
	\label{fig:ecc000maps}
	\end{center}
	\end{figure}

In addition to the inclination of the system, the orbital lightcurve of an eccentric planet depends also on the orientation of the orbit relative to the observer.
Here we consider only observers in the plane of the orbit that see the planet in superior conjunction (full dayside) either at periastron or apoastron. 
In these configurations, a transit occurs at inferior conjunction and an eclipse at superior conjunction although these events are not included here. 
As shown by \citet{Selsis2011, Selsis2013} and \citet{Maurin2012}, lightcurves are only moderately sensitive to the inclination between 90$^{\circ}$ (transit geometry) and 60$^{\circ}$ (the median value). 
At an inclination of 0$^{\circ}$ (polar observer), the phase angle is constant and the lightcurve is only modulated by the change of orbital distance. 
In general, the obliquity also introduces a seasonal modulation \citep{Gaidos2004} but all our cases have a null obliquity. 
In this article we only describe observables obtained for the $\Loo$ case. 
The idea is more to illustrate the connection between climate and observables rather than to prepare the characterization of Earth-like planets around Sun-like stars, as no forthcoming instrument will be able to provide such data. 
JWST, on the other hand, may be able to obtain (at a large observing-time cost) some data for terrestrial habitable-zone planets around cool host-dwarfs \citep{Cowan2015, Triaud2013, Belu2013}. 
However, as before, without accounting for the actual spectral distribution of the host, which will be the object of a next study, the generated signatures of planets around M stars and brown dwarfs would not be realistic.

\subsection{Circular case}

Figure~\ref{fig:ecc000maps} has been obtained for one orbit and a sub-observer point initially located arbitrarily on the equator. 
Each horizontal line of the plot represents the planet as seen by a distant observer at different orbital phases in a spectral band of the model (only the bands exhibiting a significant flux are shown, hence the gap between 2 and 6~$\mu$m). 
On the left part, the outgoing flux is shown at the spatial resolution of the GCM, while on the right part, the flux is a disk-average. 
In other words, a line on the right panel is a \textit{phase curve}, while a column is an instantaneous disk-averaged spectrum. 
To obtain the flux measured per band at a distance $d$ the value given must be diluted by a factor $(\Rearth/d)^2$. 
Beside the fact that it provides both spectral and photometric information, this type of figure is a useful tool to analyse the energy budget of the planet: we can see where and at what wavelengths the planet receives its energy and cools to space. 
The top panel of Figure~\ref{fig:phasesIR} shows for the circular case the thermal phase curves and their variability, in bands centered on 6.7~$\mu$m (H$_2$O absorption), 11~$\mu$m (window) and  15~$\mu$m (CO$_2$ absorption). The phase curve at 0.77~$\mu$m (reflected light) and its variability are shown in Figure \ref{fig:phasesVI} (black curve for the circular case). 
The long- and shortwave phase curves shown in Figure~\ref{fig:ecc000maps}, \ref{fig:phasesIR} and \ref{fig:phasesVI} present several noticeable features: \\

\noindent -  The central region of the dayside, that extends about $40^{\circ}$ from the substellar point, appears dark in the infrared due to clouds and humidity associated with the massive updraft. 
At short wavelengths, this substellar cloudy area is on the contrary the most reflective region due to the scattering by clouds (see Figure~\ref{fig:ecc000maps}). \\

\noindent - As the modeled planet has a null obliquity and is locked in a 1:1 spin-orbit resonance, external forcing is constant at any given point on the planet. 
The 3D structure of the planet that influences the emerging fluxes is not, however, constant. 
Clouds in particular produce stochastic meteorological phenomena that induce variations in the observables. 
The amplitude of this variability is shown in Figure\ref{fig:phasesVI} for one visible band and in Figure~\ref{fig:phasesIR} for 3 infrared bands probing different levels: the surface (11~$\mu$m),  the lower and mid-atmosphere (6.7~$\mu$m, H$_2$O band), and upper layers (15~$\mu$m, CO$_2$ band). \\

\noindent - The CO$_2$ absorption band at $15~\mu$m band is the most noticeable spectral feature but it exhibits no phase-variation. 
This is due to the fact that the atmospheric layers emitting to space in this band  ($\sim$50-80~mbar) are efficiently homogenized by circulation. 
This can be seen in Fig~\ref{profiles_T_h2oi_h2ov_ecc000_dayside}: the mean day and night thermal profiles are nearly identical in the range 400-20~mbar and clouds (on the dayside) do not extend at altitudes above the 100~mbar layer. \\

\noindent - Most of the cooling occurs within the 9-13~$\mu$m atmospheric window, which is transparent down to the surface in the absence of clouds, and between 16 and 24~$\mu$m, a domain affected by a significant H$_2$O absorption increasing with the wavelength. In the 9-13~$\mu$m atmospheric window, most of the emission takes place on the cloud-free ring of the dayside. 
Except for the cloudy substellar region, the brightness temperature in this window is close to the surface temperature and thus goes from an average of around 270-280~K on the dayside down to 240-250~K on the nightside. 
In this window, the orbital lightcurve therefore peaks at superior conjunction, as we can see in the top panel of Figure~\ref{fig:phasesIR}.\\

\noindent - At thermal wavelengths absorbed by water vapor (5-7 $\mu$m and above 16~$\mu$m) most of the emission takes place on the nightside, producing light curves that peak at inferior conjunction. 
The 6.7~$\mu$m and 11~$\mu$m lightcurves in the top panel of Figure~\ref{fig:phasesIR} are therefore in phase opposition. 
On the dayside, the large columns of water vapor result in a high altitude, and therefore cold, 6.7~$\mu$m photosphere (around 200~mbar, 240~K). On the nightside, the 6.7~$\mu$m photosphere is the surface due to the low humidity (around 270-280 K). \\

\noindent - \citet{Yang2013} and \citet{Illeana2013} noted that the wavelength-integrated emitted and scattered light have opposed phase variations. 
We also find this behavior because the emission at $\lambda > 16~\mu$m represents the larger part of the bolometric cooling, as we can see in Figure~\ref{fig:ecc000maps}. 
However, the emission in the 9-13~$\mu$m atmospheric band carries a significant fraction of the bolometric emission and does peak at superior conjunction as reflected light does. 
A broadband that includes both spectral regions (9-13 and 16-25~$\mu$m) therefore mixes two opposite phase variations while distinguishing between these two domains of the thermal emission could provide strong constrains on the nature of the atmosphere.
Although this should be explored further, this opposite phase variations between the  9-13 and 16-25~$\mu$m intervals may be a strong sign of both synchronization and the massive presence of water. 
Adequate filters at thermal wavelength could be used to exploit this signature while maximizing the S/N.\\

\subsection{Eccentric cases}

\begin{figure}[htbp!]
	\begin{center}
	\includegraphics[width=\linewidth]{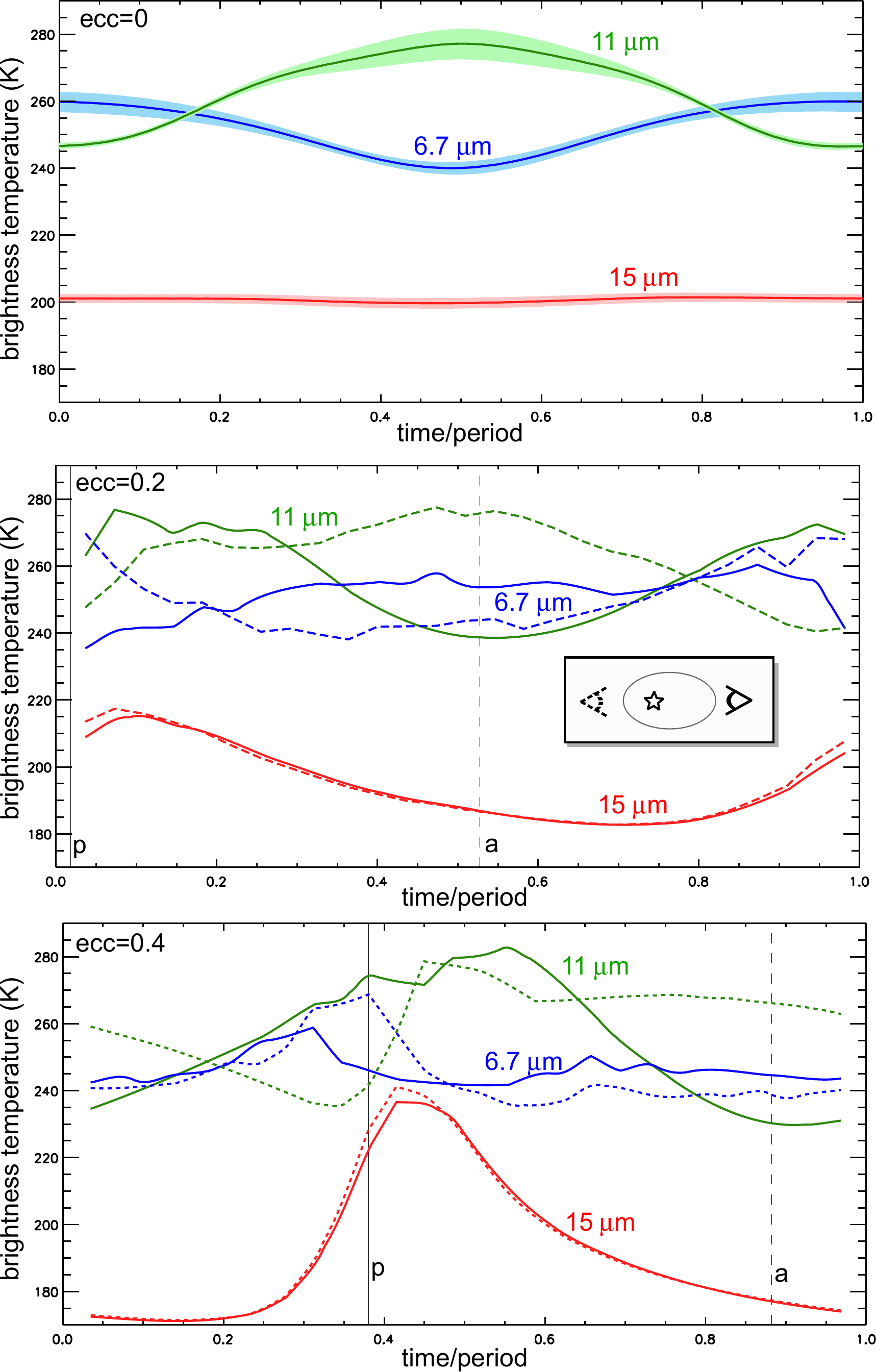}
	\caption{Thermal phase curves in the 6.7, 11 and 15~$\mu$m bands obtained for one orbit, for three eccentricities and two observation geometries. In the circular case, lightcurves are centered on the superior conjonction (dayside in view) and are time-averaged while the grey area is the 1-$\sigma$ variability due to meteorology. For a given eccentricity, the same orbit has been used for both observation geometries. The periastron (p) and apoastron (a) are indicated by vertical lines.}
	\label{fig:phasesIR}
	\end{center}
	\end{figure}

\begin{figure}[htbp!]
	\begin{center}
	\includegraphics[width=\linewidth]{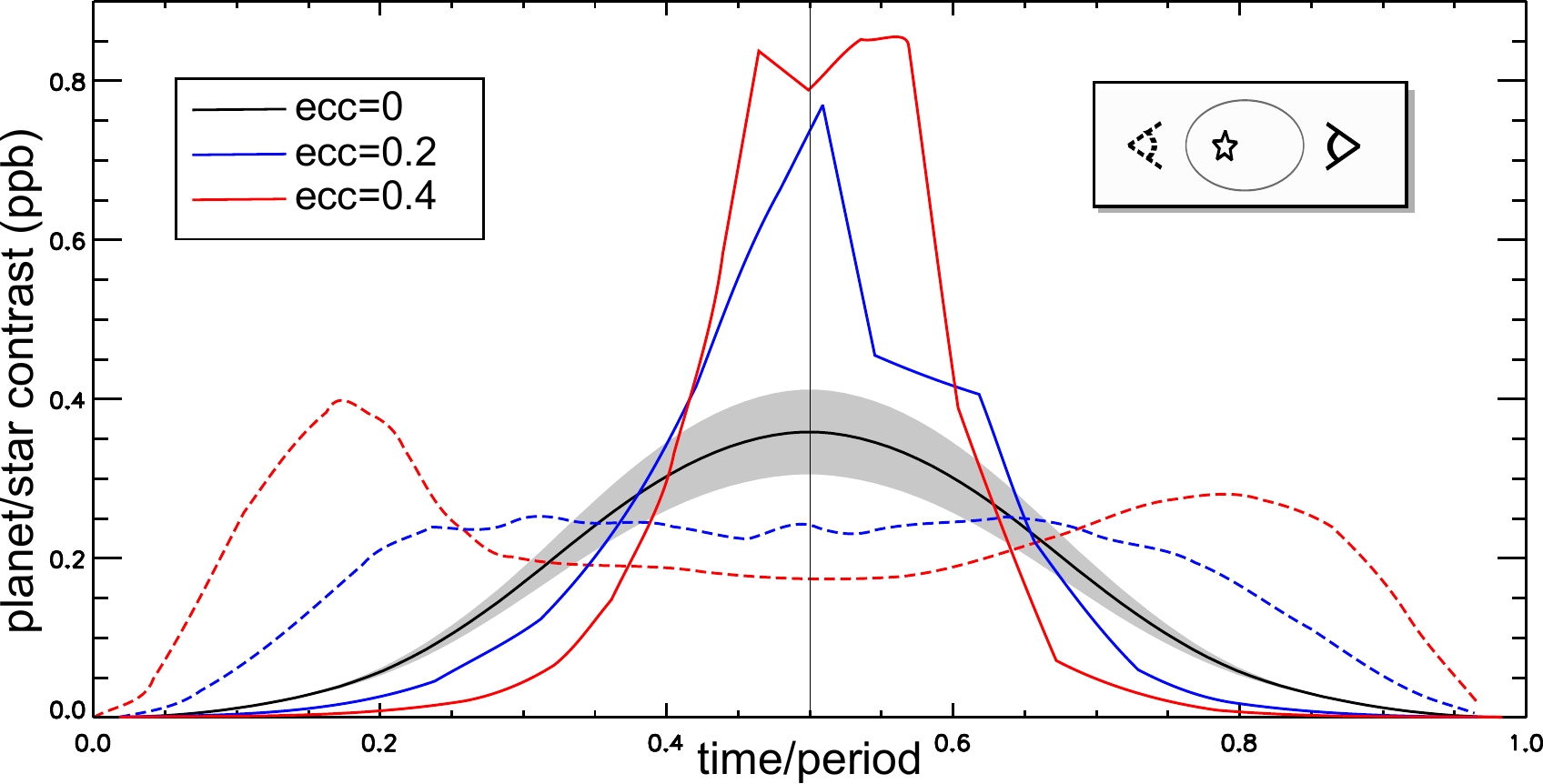}
	\caption{Reflected-light phase curves in the 0.77~$\mu$m band obtained for one orbit, for three eccentricities and two observation geometries. In the circular case, lightcurves are centered on the superior conjonction (dayside in view) and are time-averaged while the grey area is the 1-$\sigma$ variability due to meteorology. For a given eccentricity, the same orbit has been used for both observation geometries but, unlike Figure~\ref{fig:phasesIR}, they are shown here with half a period offset in order to be both centered on the superior conjunction (vertical line).}
	\label{fig:phasesVI}
	\end{center}
	\end{figure}

\begin{figure*}[htbp!]
	\begin{center}
	\includegraphics[width=\linewidth]{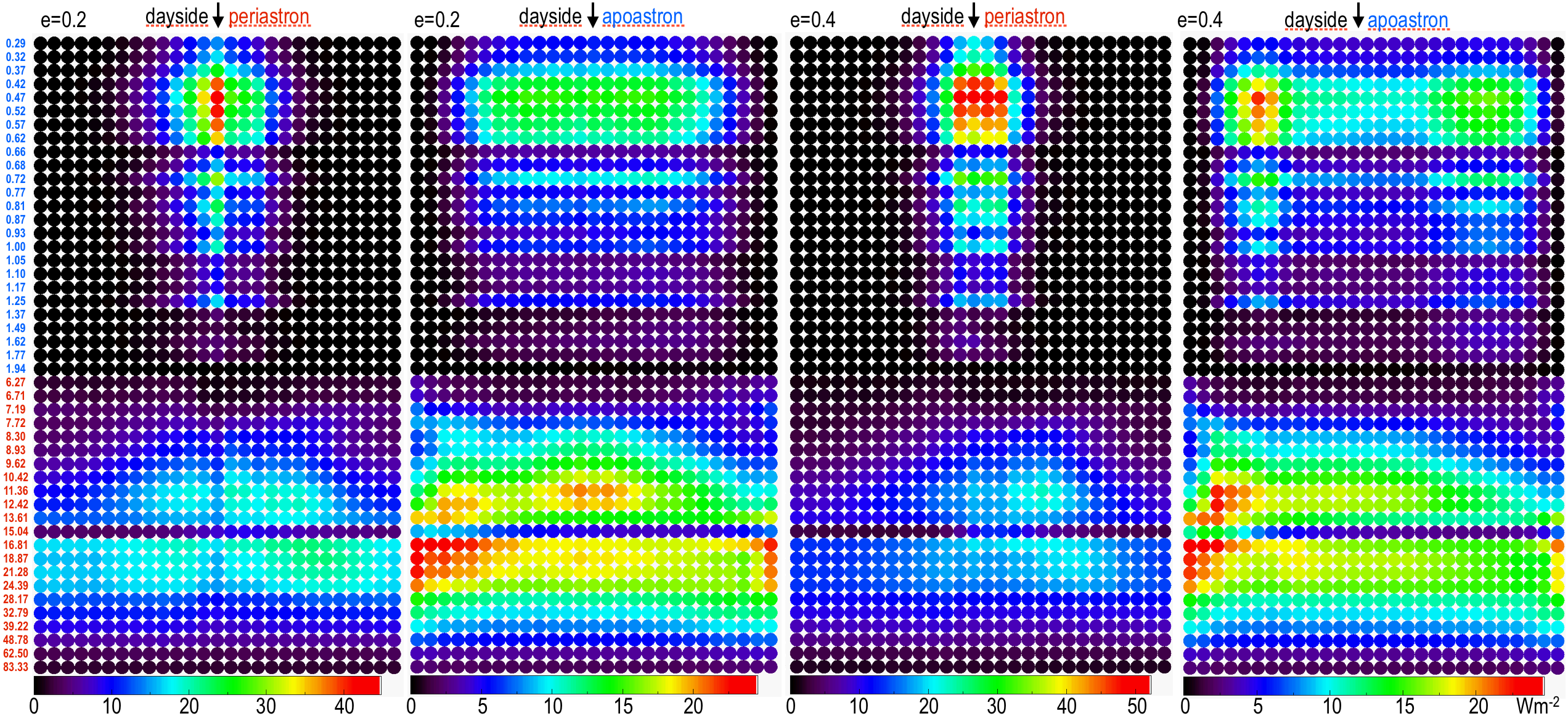}
	\caption{This figure is similar to the right panel of Figure~\ref{fig:ecc000maps} but for eccentric cases. Two observation geometries are shown for each eccentricity: the arrows on top indicate whether the dayside is observed at apoastron or periastron. The (different) color scales indicate the spatially-unresolved flux.}
	\label{fig:e020e040}
	\end{center}
	\end{figure*}

The synthetic observables obtained for $\Loo$ and $e=0.2$ and $e=0.4$ are given in Figure~\ref{fig:phasesIR} (infrared lightcurves), \ref{fig:phasesVI} (reflection lightcurves) and \ref{fig:e020e040} (full orbital spectro-photometry). 
Because the maximum eccentricity of an observed possibly-rocky planet is only $0.27$ (see Section \ref{intro}), we choose to show our cases up to an eccentricity of $e=0.4$.
Several features can be noted: \\

\noindent - The emission in the 15~$\mu$m CO$_2$ band remains phase-independent due to efficient heat redistribution at altitudes above the 100~mb layer. 
As a consequence, the two observation geometries produce the same lightcurve, which variations are only in response to the change of orbital distance (with some lag due to the inertia of the system).
This is particularly noticeable in Figure~\ref{fig:phasesIR}. \\

	\begin{figure}[htbp!]
	\begin{center}
	\includegraphics[width=\linewidth]{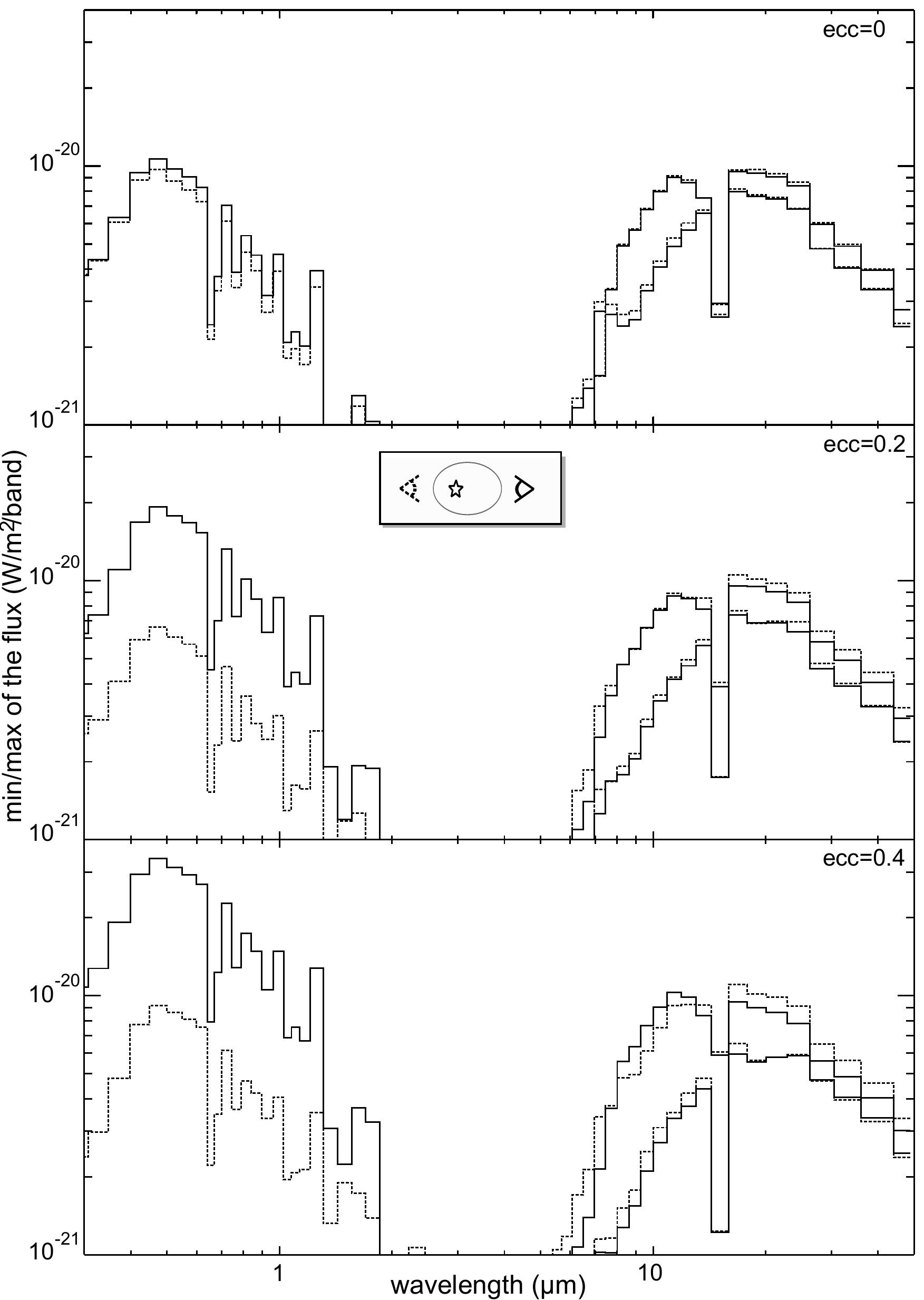}
	\caption{Spectral variability. These graphs show, for $\Ls = \Lsun$ and e = 0, 0.2 and 0.4, the maximum and minimum of the flux in each band received during the one orbit. Flux is given at 10~pc. For the reflected light, the minimum is zero as the observer is in the plane of the orbit (inclination = 90$\deg$). Superior conjunction occurs at periastron for solid lines and at apoastron for dashed lines. For the circular case (top), solid and dashed lines indicate two arbitrary observer direction separated by 180$\deg$ and, in this case, the difference between the two curves is only due to stochastic meteorological variations.}
	\label{specVIIR_var}
	\end{center}
	\end{figure}

\noindent - The phase opposition that exists in the circular case between the scattered light and the emission in the 8-12~$\mu$m atmospheric window on the one hand and the emission in the 16-25~$\mu$m water vapor window on the other hand disappears at $e=0.4$ and is hardly seen at $e=0.2$ except when the superior conjunction appears at apoastron (second panel in Figure~\ref{fig:e020e040}). 
This can be explained by the strong periastron warming that generates a cloudiness that covers a larger fraction of the sunlit hemisphere and persists on the nightside. 
These clouds hence decrease the infrared cooling in the atmospheric window on the dayside. \\

\noindent - Another effect of this large cloud coverage is to spread the observed scattered light over a broader range of phase angles, with no peak at superior conjunction when it occurs at apoastron. This can be seen on Figure~\ref{fig:phasesVI} and \ref{fig:e020e040}. \\

\begin{table}[htbp]
\begin{center}
\caption{Bond albedo computed and "measured" the $\Loo$ cases. $A_{\textrm{obs}}$ is the Bond albedo estimated by a distant observer with broadband photometry (see text). The range of values results from changing the observing geometry as well as meteorology. For the polar observer, we assume that the radius of the planet is unknown. }
\vspace{0.1cm}
\begin{tabular}{lccc}
\hline
$e$ & 0.0 & 0.2 & 0.4 \\
\hline
$A_{\textrm{Bond}}$ & 0.25 & 0.295 & 0.35 \\
\hline
$A_{\textrm{obs}}$ & & &  \\
\multicolumn{4}{l}{Equatorial observer ($i=90^{\circ}$)} \\
\small{SW} & 0.26-0.29 &  0.29-0.34 & 0.34-0.43 \\
\small{LW} & 0.27-0.29 & 0.28-0.35 &  0.34-0.42 \\
\small{LW/SW} & 0.26-0.29 & 0.29-0.35 & 0.34-0.42 \\
\multicolumn{4}{l}{Polar observer ($i=0^{\circ}$)} \\
\small{LW/SW} & 0.22-0.23 & 0.26-0.27 & 0.32-0.34 \\
\hline
\end{tabular} 
\label{tab:alb} 
\end{center}
\end{table}

\subsection{Variation spectrum}

The variation spectrum, as defined in Selsis et al. \citep{Selsis2011}, is the peak amplitude of the phase curve as a function of wavelength. 
Figure~\ref{specVIIR_var} shows the minimum and maximum that can be observed anytime during a complete orbit for each band of the GCM. 
Two opposite geometries are presented, for $e=0, 0.2$ and $0.4$ and $\Ls=\Lsun$. 
The difference between the maximum and the minimum gives the amplitude spectrum at the resolution of the GCM. 
As we only consider cases with a $90^{\circ}$ inclination, the reflected flux is always null at inferior conjunction (transit). 
The reflected variation spectrum naturally depends on the eccentricity as the closest the planet approaches its hot star, the highest its reflected brightness (for a given albedo). 
The thermal variation spectrum shows, on the other hand, an overall minor dependency on the eccentricity. 
The amplitude of the differences in thermal flux does increase slightly with eccentricity but the differences between one eccentricity to another are not significantly higher than those induced by stochastic meteorological changes from one orbit to another. 
One exception is the amplitude of the variations in the $15~\mu$m band, which increases notably with the eccentricity, from basically no variations at $e=0$ to a factor of 2 variation at $e=0.2$ and a factor of 5 at $e=0.4$.

\subsection{Radiative budget}

An observer that would have the ability to measure the orbit-averaged broadband emission at short (SW: 0.25-2.0~$\mu$m) and/or long (LW: 6-35~$\mu$m) wavelengths, would be able to estimate the Bond albedo. 
This could be done with either of these two values if knowing the planetary radius and the incoming stellar flux, or from the LW/SW ratio without any other information. 
These estimates would however be biased by the observing geometry and emission/reflection anisotropies. 
We therefore tested how these measured values of the Bond albedo depart from its actual value. 
The actual Bond albedo calculated from the simulations as well as those that can be estimated by an observer are given in the Table~\ref{tab:alb}. 
We can see that the error is typically of the order of 10-20\%, with a slight tendency to overestimate the Bond albedo. 
This is due to the fact that the reflected light is more focused into the orbital plane than the thermal emission, which is re-distributed by circulation. A polar observer may on the contrary underestimate the Bond albedo.


\section{Discussion}
\label{Discussion}

We studied here the influence of the duration of the orbital period and eccentricity on the climate of tidally locked ocean covered planets orbiting objects of different luminosities.
We chose here to do a parametric study with most things equal between our simulations: we considered synchronized planets and we did not take into account the spectral differences between a Sun-like star and a lower luminosity object.

Low mass stars have a redder spectrum and this changes the albedo of the ice and snow \citep{JoshiHaberle2012, vonParis2013}.
Taking this phenomenon into account would radically change the ice-albedo feedback, so that for redder objects it would not be able to drive the planets into a glaciation state (such as what we showed here for planets with high eccentricities around a $\Ls = \Lot$ star). 
\citet{Godolt2015} showed that for the same incoming flux, the planets orbiting redder objects would be hotter, and generally would have different climate states than those orbiting a $\Loo$ star \citep[see also][]{Shields2014}.

Our results would also be affected had we taken into account a realistic oceanic circulation. 
Oceanic circulation would have the effect of homogenizing the climate and facilitate the existence of habitable states.
\citet{Godolt2015} shows that taking into account oceanic circulation allows them to find surface habitable conditions for planets orbiting F-type stars instead of a snowball state.
Our results would also be likely to change if we had considered other types of planets such as an Earth-like planet, a land planet, a planet with a Pangea-like continent, a planet with archipelagos \citep[e.g.][]{Yang2014}. 
 
Besides, we consider here a planet whose atmosphere has a composition very similar to the Earth's composition. 
However, the atmospheric composition could be different.
For instance, there could be more CO$_2$, which would contribute to heat up the planet. 
Also, the pressure in the atmosphere could be different, which would affect the climatic response to eccentricity.
 
In this work we also made the choice of neglecting the effects of tides even though we considered high eccentricities. 
Tides would have several effects on the system considered here.
First, tides act to damp the eccentricities of planets, so that a planet with an eccentricity of 0.8 would not keep its eccentricity forever \citep[the timescale of eccentricity damping depending on many parameters such as the orbital parameters, the mass, radius and composition of the planet; e.g.][]{Mignard1979, Hut1981}.
However, a high eccentricity could be maintained if the planet is part of a multi-planet system. 
Indeed the gravitational interactions between the planets can contribute in exciting the eccentricity of the planet on long timescales. 
Second, such high eccentricities can be responsible for tidal heating in the interior of the planet \citep{Jackson2008, Barnes2009a, Bolmont2014sf2a}. 
This tidal heating is all the more important as we consider planets orbiting low luminosity objects. 
Indeed, a planet around a less luminous star and receiving the same average flux as Earth is much closer to its host star and thus susceptible to a higher tidal heating.
This effect should be studied, as tidal heating could contribute significantly to the energy budget of the planet \citep[and create ``tidal Venuses'' as discussed in][]{Barnes2009a}.
Third, tides also influence the rotation state of the planet. 
We considered here tidally locked planets, but planets on eccentric orbits are more likely to have a pseudo-synchronous rotation \citep[synchronization around periastron, see][]{Hut1981} or to be in spin-orbit resonance \citep[the higher the eccentricity, the higher the order of the resonance, see][]{Makarov2013}.
The pseudo-synchronous rotation and the spin-orbit resonance are faster than the synchronous rotation, so the wind will be even more efficient to redistribute the heat on the planet. 
Finally, even for a circular orbit, atmospheric tides can drive a planet out of synchronization \citep{Correia2003, Leconte2015}.

In this work, we focused only on the presence of surface liquid water without having to conclude about the actual potential of the planets to be appropriate environments for the apparition of life.
Indeed, the question of the apparition of life on water worlds is still open. 
In order for life to appear and evolve, water must be in contact with the building bricks of life, such as phosphorus, sulfur, iron, magnesium and nitrogen.
This is possible if the ocean is in direct contact with the planet's silicate mantle.
However, \citet{Sotin2007} found that ocean planets are likely to have a high-pressure ice layer between the liquid water ocean and the silicate mantle. 
Therefore, \citet{Lammer2009} sorted the ocean planets as class IV habitat, the lowest class of potentially habitable worlds \citep[see also][]{Forget2013}.
However, \citet{Leger2004} and \citet{Forget2013} both highlighted the possibility to enrich the ocean with minerals from meteoritic impacts, showing that these planetary objects should not be discarded as potential life habitats.
Besides, some other mechanisms were put forward to bring minerals to the liquid water layer, such as solid convection \citep[e.g., for example the presence of $^{36}$Ar at the surface of Titan could be explained by solid convection within the subsurface high pressure ice layer;][]{Niemann2010, Tobie2006}.
Finally \citet{Alibert2015} showed that planets with a small mass fraction of water ($\sim 1-2$\%, depending on the mass of the planet) have more probability to be habitable due to the absence of a high pressure ice layer.

\section{Conclusions}
\label{Conclusions}

Thanks to climate simulations of synchronous ocean covered planets, we are able to assess their potential habitability in the sense of their capacity to host a liquid water ocean at their surface. 

We investigated whether the mean flux approximation \citep{WilliamsPollard2002a} is correct to assess the habitability of eccentric planets orbiting stars of different luminosity.
In order to do so, we considered planets receiving on average the same flux as Earth and modeled their climates for different eccentricities and different orbital periods (used here as a proxy for the three different star luminosities we considered: $\Loo$, $\Lot$, $\Lof$). 

We found that tidally locked water worlds can sustain surface liquid water only on the dayside.
For all luminosities and only small eccentricities, all the planets considered can always sustain surface liquid water on the dayside.
Planets orbiting the less luminous objects considered here ($\Lof$) can always sustain surface liquid water whatever is the eccentricity. 

\begin{figure}[htbp!]
	\begin{center}
	\includegraphics[width=9cm]{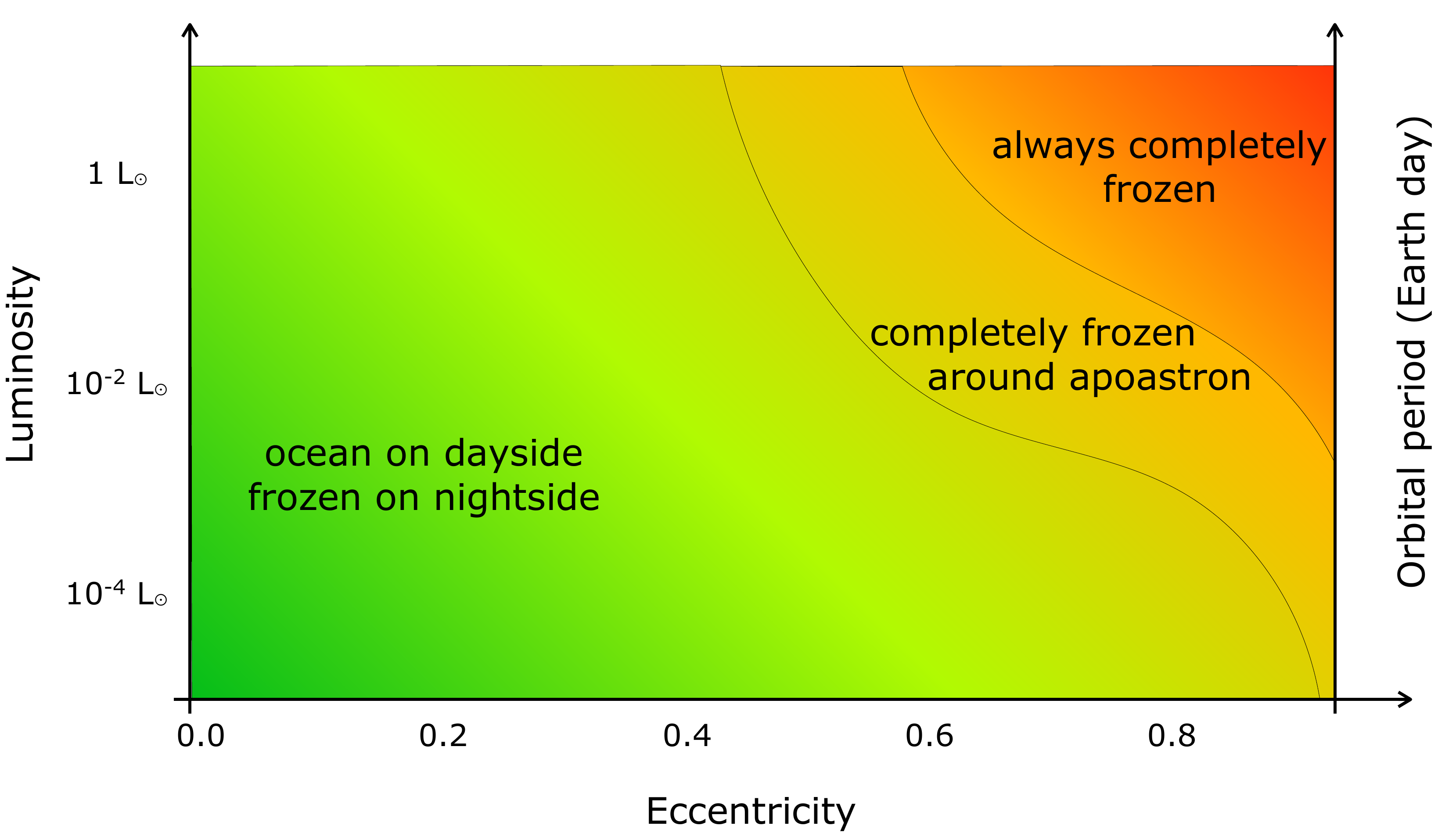}
	\caption{Liquid water coverage map of tidally locked ocean covered planets orbiting stars of different luminosities and with different orbital eccentricities.}
	\label{conclusion_graph}
	\end{center}
	\end{figure}

However, planets on high eccentricity orbits around luminous objects can only sustain surface liquid water around periastron.
This is the case for planets orbiting a $\Loo$ star with an eccentricity higher than 0.6, planets orbiting a $\Lot$ star with an eccentricity higher than 0.8 and planets orbiting a $\Lof$ star with an eccentricity higher than 0.9.
For exemple, we find that a planet orbiting a $\Lot$ star with an eccentricity of 0.9 cannot have surface liquid water, it always has an ice layer covering the whole planet. 
Besides, for planets with high eccentricities, the dayside temperature variations over a period of 365 days ($\Loo$) to 4 days ($\Lof$) can be huge (up to 100~K). 
This could have detrimental consequences for eventual life forms.
Figure \ref{conclusion_graph} summarizes our results.

For the planets considered here, i.e. water worlds planets for which the surface is treated as an infinite water source, tidally locked and with no obliquity, we found that the higher the eccentricity of the planet or the higher the luminosity of the star, the less reliable the mean flux approximation.
These results are not in agreement with the work of \citet{WilliamsPollard2002a}, which assessed that the mean flux approximation is valid for all eccentricities.
When considering $\Ls = \Loo$, we draw similar conclusions to those of \citet{Linsenmeier2015}, who pointed out that planets on eccentric orbits around a Sun-like star can be frozen during part of the year.

We explored the parameter space of our model by changing the thermal inertia of the ocean and the maximum ice layer thickness.
Changing those parameters in the model does not change significantly the conclusions above.

We will address in a future work the limits discussed in the previous section, in particular the spectral difference between low mass stars and the Sun, the distribution of continents and tidal effects.


\begin{acknowledgements}  

This work was supported by the Fonds de la Recherche Scientifique-FNRS under Grant No. T.0029.13 (``ExtraOrDynHa'' research project).
Computational resources have been provided by the Consortium des \'Equipements de Calcul Intensif (C\'ECI), funded by the Fonds de la Recherche Scientifique de Belgique (F.R.S.-FNRS) under Grant No.2.5020.11.
The authors would like to thank Ehouarn Millour and Aymeric Spiga for their help in this work and Stephen Kane for the Habitable Zone Gallery.

\end{acknowledgements}



\begin{thebibliography}{77}
\expandafter\ifx\csname natexlab\endcsname\relax\def\natexlab#1{#1}\fi

\bibitem[{{Alibert}(2015)}]{Alibert2015}
{Alibert}, Y. 2015, Origins of Life and Evolution of the Biosphere, 45, 319

\bibitem[{{Anglada-Escud{\'e}} {et~al.}(2012){Anglada-Escud{\'e}}, {Arriagada},
  {Vogt}, {Rivera}, {Butler}, {Crane}, {Shectman}, {Thompson}, {Minniti},
  {Haghighipour}, {Carter}, {Tinney}, {Wittenmyer}, {Bailey}, {O'Toole},
  {Jones}, \& {Jenkins}}]{Anglada-Escude2012}
{Anglada-Escud{\'e}}, G., {Arriagada}, P., {Vogt}, S.~S., {et~al.} 2012, \apjl,
  751, L16

\bibitem[{{Bailey} {et~al.}(2009){Bailey}, {Butler}, {Tinney}, {Jones},
  {O'Toole}, {Carter}, \& {Marcy}}]{Bailey2009}
{Bailey}, J., {Butler}, R.~P., {Tinney}, C.~G., {et~al.} 2009, \apj, 690, 743

\bibitem[{{Barnes} {et~al.}(2009){Barnes}, {Jackson}, {Greenberg}, \&
  {Raymond}}]{Barnes2009a}
{Barnes}, R., {Jackson}, B., {Greenberg}, R., \& {Raymond}, S.~N. 2009, ApJ
  Lett., 700, L30

\bibitem[{{Barnes} {et~al.}(2008){Barnes}, {Raymond}, {Jackson}, \&
  {Greenberg}}]{Barnes2008}
{Barnes}, R., {Raymond}, S.~N., {Jackson}, B., \& {Greenberg}, R. 2008,
  Astrobiology, 8, 557

\bibitem[{{Belu} {et~al.}(2013){Belu}, {Selsis}, {Raymond}, {Pall{\'e}},
  {Street}, {Sahu}, {von Braun}, {Bolmont}, {Figueira}, {Anupama}, \&
  {Ribas}}]{Belu2013}
{Belu}, A.~R., {Selsis}, F., {Raymond}, S.~N., {et~al.} 2013, ApJ, 768, 125

\bibitem[{{Bolmont} {et~al.}(2014){Bolmont}, {Raymond}, \&
  {Selsis}}]{Bolmont2014sf2a}
{Bolmont}, E., {Raymond}, S.~N., \& {Selsis}, F. 2014, in SF2A-2014:
  Proceedings of the Annual meeting of the French Society of Astronomy and
  Astrophysics, ed. J.~{Ballet}, F.~{Martins}, F.~{Bournaud}, R.~{Monier}, \&
  C.~{Reyl{\'e}}, 63--68

\bibitem[{{Borucki} {et~al.}(2011){Borucki}, {Koch}, {Basri}, {Batalha},
  {Brown}, {Bryson}, {Caldwell}, {Christensen-Dalsgaard}, {Cochran}, {DeVore},
  {Dunham}, {Gautier}, {Geary}, {Gilliland}, {Gould}, {Howell}, {Jenkins},
  {Latham}, {Lissauer}, {Marcy}, {Rowe}, {Sasselov}, {Boss}, {Charbonneau},
  {Ciardi}, {Doyle}, {Dupree}, {Ford}, {Fortney}, {Holman}, {Seager},
  {Steffen}, {Tarter}, {Welsh}, {Allen}, {Buchhave}, {Christiansen}, {Clarke},
  {Das}, {D{\'e}sert}, {Endl}, {Fabrycky}, {Fressin}, {Haas}, {Horch},
  {Howard}, {Isaacson}, {Kjeldsen}, {Kolodziejczak}, {Kulesa}, {Li}, {Lucas},
  {Machalek}, {McCarthy}, {MacQueen}, {Meibom}, {Miquel}, {Prsa}, {Quinn},
  {Quintana}, {Ragozzine}, {Sherry}, {Shporer}, {Tenenbaum}, {Torres},
  {Twicken}, {Van Cleve}, {Walkowicz}, {Witteborn}, \& {Still}}]{Borucki2011}
{Borucki}, W.~J., {Koch}, D.~G., {Basri}, G., {et~al.} 2011, ApJ, 736, 19

\bibitem[{{Butcher} {et~al.}(1992){Butcher}, {Charlson}, {Orians}, \&
  {Wolfe}}]{Butcher1992}
{Butcher}, S.~S., {Charlson}, R.~J., {Orians}, G.~H., \& {Wolfe}, G.~V. 1992,
  {Global Biogeochemical Cycles} (Academic Press, London)

\bibitem[{{Carone} {et~al.}(2015){Carone}, {Keppens}, \& {Decin}}]{Carone2015}
{Carone}, L., {Keppens}, R., \& {Decin}, L. 2015, \mnras, 453, 2412

\bibitem[{{Clough} {et~al.}(1989){Clough}, {Kneizys}, \& {Davies}}]{Clough1989}
{Clough}, S.~A., {Kneizys}, F.~X., \& {Davies}, R.~W. 1989, Atmospheric
  Research, 23, 229

\bibitem[{{Correia} {et~al.}(2003){Correia}, {Laskar}, \& {de
  Surgy}}]{Correia2003}
{Correia}, A.~C.~M., {Laskar}, J., \& {de Surgy}, O.~N. 2003, \icarus, 163, 1

\bibitem[{{Cowan} \& {Agol}(2011)}]{CowanAgol2011}
{Cowan}, N.~B. \& {Agol}, E. 2011, \apj, 726, 82

\bibitem[{{Cowan} {et~al.}(2015){Cowan}, {Greene}, {Angerhausen}, {Batalha},
  {Clampin}, {Col{\'o}n}, {Crossfield}, {Fortney}, {Gaudi}, {Harrington},
  {Iro}, {Lillie}, {Linsky}, {Lopez-Morales}, {Mandell}, {Stevenson}, \&
  {ExoPAG SAG-10}}]{Cowan2015}
{Cowan}, N.~B., {Greene}, T., {Angerhausen}, D., {et~al.} 2015, \pasp, 127, 311

\bibitem[{{Dressing} {et~al.}(2010){Dressing}, {Spiegel}, {Scharf}, {Menou}, \&
  {Raymond}}]{Dressing2010}
{Dressing}, C.~D., {Spiegel}, D.~S., {Scharf}, C.~A., {Menou}, K., \&
  {Raymond}, S.~N. 2010, \apj, 721, 1295

\bibitem[{{Ehrenreich} {et~al.}(2015){Ehrenreich}, {Bourrier}, {Wheatley},
  {Lecavelier des Etangs}, {H{\'e}brard}, {Udry}, {Bonfils}, {Delfosse},
  {D{\'e}sert}, {Sing}, \& {Vidal-Madjar}}]{Ehrenreich2015}
{Ehrenreich}, D., {Bourrier}, V., {Wheatley}, P.~J., {et~al.} 2015, \nat, 522,
  459

\bibitem[{{Forget} {et~al.}(2013){Forget}, {Wordsworth}, {Millour},
  {Madeleine}, {Kerber}, {Leconte}, {Marcq}, \& {Haberle}}]{Forget2013}
{Forget}, F., {Wordsworth}, R., {Millour}, E., {et~al.} 2013, \icarus, 222, 81

\bibitem[{{Gaidos} \& {Williams}(2004)}]{Gaidos2004}
{Gaidos}, E. \& {Williams}, D.~M. 2004, \na, 10, 67

\bibitem[{{Godolt} {et~al.}(2015){Godolt}, {Grenfell}, {Hamann-Reinus},
  {Kitzmann}, {Kunze}, {Langematz}, {von Paris}, {Patzer}, {Rauer}, \&
  {Stracke}}]{Godolt2015}
{Godolt}, M., {Grenfell}, J.~L., {Hamann-Reinus}, A., {et~al.} 2015, \planss,
  111, 62

\bibitem[{{G{\'o}mez-Leal}(2013)}]{Illeana2013}
{G{\'o}mez-Leal}, I. 2013, PhD thesis, University of Bordeaux, France

\bibitem[{{Hourdin} {et~al.}(2006){Hourdin}, {Musat}, {Bony}, {Braconnot},
  {Codron}, {Dufresne}, {Fairhead}, {Filiberti}, {Friedlingstein}, {Grandpeix},
  {Krinner}, {LeVan}, {Li}, \& {Lott}}]{Hourdin2006}
{Hourdin}, F., {Musat}, I., {Bony}, S., {et~al.} 2006, Climate Dynamics, 27,
  787

\bibitem[{{Hut}(1981)}]{Hut1981}
{Hut}, P. 1981, A \& A, 99, 126

\bibitem[{{Iro} \& {Deming}(2010)}]{IroDeming2010}
{Iro}, N. \& {Deming}, L.~D. 2010, \apj, 712, 218

\bibitem[{{Jackson} {et~al.}(2008){Jackson}, {Greenberg}, \&
  {Barnes}}]{Jackson2008}
{Jackson}, B., {Greenberg}, R., \& {Barnes}, R. 2008, ApJ, 681, 1631

\bibitem[{{Joshi} \& {Haberle}(2012)}]{JoshiHaberle2012}
{Joshi}, M.~M. \& {Haberle}, R.~M. 2012, Astrobiology, 12, 3

\bibitem[{{Kaltenegger} {et~al.}(2013){Kaltenegger}, {Sasselov}, \&
  {Rugheimer}}]{Kaltenegger2013}
{Kaltenegger}, L., {Sasselov}, D., \& {Rugheimer}, S. 2013, \apjl, 775, L47

\bibitem[{{Kane} \& {Gelino}(2011)}]{KaneGelino2011}
{Kane}, S.~R. \& {Gelino}, D.~M. 2011, \apj, 741, 52

\bibitem[{{Kane} \& {Gelino}(2012)}]{KaneGelino2012}
{Kane}, S.~R. \& {Gelino}, D.~M. 2012, \pasp, 124, 323

\bibitem[{{Kasting} {et~al.}(1993){Kasting}, {Whitmire}, \&
  {Reynolds}}]{Kasting1993}
{Kasting}, J.~F., {Whitmire}, D.~P., \& {Reynolds}, R.~T. 1993, Icarus, 101,
  108

\bibitem[{{Kataria} {et~al.}(2013){Kataria}, {Showman}, {Lewis}, {Fortney},
  {Marley}, \& {Freedman}}]{Kataria2013}
{Kataria}, T., {Showman}, A.~P., {Lewis}, N.~K., {et~al.} 2013, \apj, 767, 76

\bibitem[{{Kopparapu}(2013)}]{Kopparapu2013}
{Kopparapu}, R.~K. 2013, ApJL, 767, L8

\bibitem[{{Kopparapu} {et~al.}(2014){Kopparapu}, {Ramirez}, {SchottelKotte},
  {Kasting}, {Domagal-Goldman}, \& {Eymet}}]{Kopparapu2014}
{Kopparapu}, R.~K., {Ramirez}, R.~M., {SchottelKotte}, J., {et~al.} 2014,
  \apjl, 787, L29

\bibitem[{{Kuchner}(2003)}]{Kuchner2003}
{Kuchner}, M.~J. 2003, \apjl, 596, L105

\bibitem[{{Kulikov} {et~al.}(2007){Kulikov}, {Lammer}, {Lichtenegger}, {Penz},
  {Breuer}, {Spohn}, {Lundin}, \& {Biernat}}]{Kulikov2007}
{Kulikov}, Y.~N., {Lammer}, H., {Lichtenegger}, H.~I.~M., {et~al.} 2007, \ssr,
  129, 207

\bibitem[{{Lammer} {et~al.}(2009){Lammer}, {Bredeh{\"o}ft}, {Coustenis},
  {Khodachenko}, {Kaltenegger}, {Grasset}, {Prieur}, {Raulin}, {Ehrenfreund},
  {Yamauchi}, {Wahlund}, {Grie{\ss}meier}, {Stangl}, {Cockell}, {Kulikov},
  {Grenfell}, \& {Rauer}}]{Lammer2009}
{Lammer}, H., {Bredeh{\"o}ft}, J.~H., {Coustenis}, A., {et~al.} 2009, \aapr,
  17, 181

\bibitem[{{Lammer} {et~al.}(2003){Lammer}, {Selsis}, {Ribas}, {Guinan},
  {Bauer}, \& {Weiss}}]{Lammer2003}
{Lammer}, H., {Selsis}, F., {Ribas}, I., {et~al.} 2003, ApJl, 598, L121

\bibitem[{{Laughlin} {et~al.}(2009){Laughlin}, {Deming}, {Langton}, {Kasen},
  {Vogt}, {Butler}, {Rivera}, \& {Meschiari}}]{Laughlin2009}
{Laughlin}, G., {Deming}, D., {Langton}, J., {et~al.} 2009, \nat, 457, 562

\bibitem[{{Leconte} {et~al.}(2013){Leconte}, {Forget}, {Charnay}, {Wordsworth},
  {Selsis}, {Millour}, \& {Spiga}}]{Leconte2013a}
{Leconte}, J., {Forget}, F., {Charnay}, B., {et~al.} 2013, \aap, 554, A69

\bibitem[{{Leconte} {et~al.}(2015){Leconte}, {Wu}, {Menou}, \&
  {Murray}}]{Leconte2015}
{Leconte}, J., {Wu}, H., {Menou}, K., \& {Murray}, N. 2015, Science, 347, 632

\bibitem[{{L{\'e}ger} {et~al.}(2004){L{\'e}ger}, {Selsis}, {Sotin}, {Guillot},
  {Despois}, {Mawet}, {Ollivier}, {Lab{\`e}que}, {Valette}, {Brachet},
  {Chazelas}, \& {Lammer}}]{Leger2004}
{L{\'e}ger}, A., {Selsis}, F., {Sotin}, C., {et~al.} 2004, \icarus, 169, 499

\bibitem[{{Lewis} {et~al.}(2013){Lewis}, {Knutson}, {Showman}, {Cowan},
  {Laughlin}, {Burrows}, {Deming}, {Crepp}, {Mighell}, {Agol}, {Bakos},
  {Charbonneau}, {D{\'e}sert}, {Fischer}, {Fortney}, {Hartman}, {Hinkley},
  {Howard}, {Johnson}, {Kao}, {Langton}, \& {Marcy}}]{Lewis2013}
{Lewis}, N.~K., {Knutson}, H.~A., {Showman}, A.~P., {et~al.} 2013, \apj, 766,
  95

\bibitem[{{Lewis} {et~al.}(2014){Lewis}, {Showman}, {Fortney}, {Knutson}, \&
  {Marley}}]{Lewis2014}
{Lewis}, N.~K., {Showman}, A.~P., {Fortney}, J.~J., {Knutson}, H.~A., \&
  {Marley}, M.~S. 2014, \apj, 795, 150

\bibitem[{{Lewis} {et~al.}(2010){Lewis}, {Showman}, {Fortney}, {Marley},
  {Freedman}, \& {Lodders}}]{Lewis2010}
{Lewis}, N.~K., {Showman}, A.~P., {Fortney}, J.~J., {et~al.} 2010, \apj, 720,
  344

\bibitem[{{Linsenmeier} {et~al.}(2015){Linsenmeier}, {Pascale}, \&
  {Lucarini}}]{Linsenmeier2015}
{Linsenmeier}, M., {Pascale}, S., \& {Lucarini}, V. 2015, \planss, 105, 43

\bibitem[{{Makarov} \& {Efroimsky}(2013)}]{Makarov2013}
{Makarov}, V.~V. \& {Efroimsky}, M. 2013, ApJ, 764, 27

\bibitem[{{Manabe} \& {Wetherald}(1967)}]{ManabeWetherald1967}
{Manabe}, S. \& {Wetherald}, R.~T. 1967, Journal of Atmospheric Sciences, 24,
  241

\bibitem[{{Maurin} {et~al.}(2012){Maurin}, {Selsis}, {Hersant}, \&
  {Belu}}]{Maurin2012}
{Maurin}, A.~S., {Selsis}, F., {Hersant}, F., \& {Belu}, A. 2012, \aap, 538,
  A95

\bibitem[{{Mignard}(1979)}]{Mignard1979}
{Mignard}, F. 1979, Moon and Planets, 20, 301

\bibitem[{{Milankovitch}(1941)}]{Milankovitch1941}
{Milankovitch}, M. 1941, {Kanon der Erdebestrahlung und seine anwendung auf das
  eiszeitenproblem} (Koniglich Serbische Akademie)

\bibitem[{{Niemann} {et~al.}(2010){Niemann}, {Atreya}, {Demick}, {Gautier},
  {Haberman}, {Harpold}, {Kasprzak}, {Lunine}, {Owen}, \&
  {Raulin}}]{Niemann2010}
{Niemann}, H.~B., {Atreya}, S.~K., {Demick}, J.~E., {et~al.} 2010, Journal of
  Geophysical Research (Planets), 115, 12006

\bibitem[{{Pierrehumbert}(2011)}]{Pierrehumbert2011}
{Pierrehumbert}, R.~T. 2011, \apjl, 726, L8

\bibitem[{{Quintana} {et~al.}(2014){Quintana}, {Barclay}, {Raymond}, {Rowe},
  {Bolmont}, {Caldwell}, {Howell}, {Kane}, {Huber}, {Crepp}, {Lissauer},
  {Ciardi}, {Coughlin}, {Everett}, {Henze}, {Horch}, {Isaacson}, {Ford},
  {Adams}, {Still}, {Hunter}, {Quarles}, \& {Selsis}}]{Quintana2014}
{Quintana}, E.~V., {Barclay}, T., {Raymond}, S.~N., {et~al.} 2014, Science,
  767, 128

\bibitem[{{Robertson} \& {Mahadevan}(2014)}]{RobertsonMahadevan2014}
{Robertson}, P. \& {Mahadevan}, S. 2014, \apjl, 793, L24

\bibitem[{{Rothman} {et~al.}(2009){Rothman}, {Gordon}, {Barbe}, {Benner},
  {Bernath}, {Birk}, {Boudon}, {Brown}, {Campargue}, {Champion}, {Chance},
  {Coudert}, {Dana}, {Devi}, {Fally}, {Flaud}, {Gamache}, {Goldman},
  {Jacquemart}, {Kleiner}, {Lacome}, {Lafferty}, {Mandin}, {Massie},
  {Mikhailenko}, {Miller}, {Moazzen-Ahmadi}, {Naumenko}, {Nikitin}, {Orphal},
  {Perevalov}, {Perrin}, {Predoi-Cross}, {Rinsland}, {Rotger}, {{\v S}ime{\v
  c}kov{\'a}}, {Smith}, {Sung}, {Tashkun}, {Tennyson}, {Toth}, {Vandaele}, \&
  {Vander Auwera}}]{Rothman2009}
{Rothman}, L.~S., {Gordon}, I.~E., {Barbe}, A., {et~al.} 2009, \jqsrt, 110, 533

\bibitem[{{Selsis} {et~al.}(2007){Selsis}, {Kasting}, {Levrard}, {Paillet},
  {Ribas}, \& {Delfosse}}]{Selsis2007}
{Selsis}, F., {Kasting}, J.~F., {Levrard}, B., {et~al.} 2007, A \& A, 476, 1373

\bibitem[{{Selsis} {et~al.}(2013){Selsis}, {Maurin}, {Hersant}, {Leconte},
  {Bolmont}, {Raymond}, \& {Delbo'}}]{Selsis2013}
{Selsis}, F., {Maurin}, A.-S., {Hersant}, F., {et~al.} 2013, A\&A, 555, A51

\bibitem[{{Selsis} {et~al.}(2011){Selsis}, {Wordsworth}, \&
  {Forget}}]{Selsis2011}
{Selsis}, F., {Wordsworth}, R.~D., \& {Forget}, F. 2011, \aap, 532, A1

\bibitem[{{Shields} {et~al.}(2014){Shields}, {Bitz}, {Meadows}, {Joshi}, \&
  {Robinson}}]{Shields2014}
{Shields}, A.~L., {Bitz}, C.~M., {Meadows}, V.~S., {Joshi}, M.~M., \&
  {Robinson}, T.~D. 2014, \apjl, 785, L9

\bibitem[{{Shields} {et~al.}(2013){Shields}, {Meadows}, {Bitz},
  {Pierrehumbert}, {Joshi}, \& {Robinson}}]{Shields2013}
{Shields}, A.~L., {Meadows}, V.~S., {Bitz}, C.~M., {et~al.} 2013, Astrobiology,
  13, 715

\bibitem[{{Showman} {et~al.}(2013){Showman}, {Fortney}, {Lewis}, \&
  {Shabram}}]{Showman2013}
{Showman}, A.~P., {Fortney}, J.~J., {Lewis}, N.~K., \& {Shabram}, M. 2013,
  \apj, 762, 24

\bibitem[{{Showman} {et~al.}(2015){Showman}, {Lewis}, \&
  {Fortney}}]{Showman2015}
{Showman}, A.~P., {Lewis}, N.~K., \& {Fortney}, J.~J. 2015, \apj, 801, 95

\bibitem[{{Showman} \& {Polvani}(2011)}]{ShowmanPolvani2011}
{Showman}, A.~P. \& {Polvani}, L.~M. 2011, \apj, 738, 71

\bibitem[{{Sotin} {et~al.}(2007){Sotin}, {Grasset}, \& {Mocquet}}]{Sotin2007}
{Sotin}, C., {Grasset}, O., \& {Mocquet}, A. 2007, \icarus, 191, 337

\bibitem[{{Spiegel} {et~al.}(2010){Spiegel}, {Burrows}, \&
  {Milsom}}]{Spiegel2010}
{Spiegel}, D.~S., {Burrows}, A., \& {Milsom}, J.~A. 2010, in AAS/Division for
  Planetary Sciences Meeting Abstracts, Vol.~42, AAS/Division for Planetary
  Sciences Meeting Abstracts 42, 27.27--+

\bibitem[{{Spiegel} {et~al.}(2008){Spiegel}, {Menou}, \&
  {Scharf}}]{Spiegel2008}
{Spiegel}, D.~S., {Menou}, K., \& {Scharf}, C.~A. 2008, \apj, 681, 1609

\bibitem[{{Tobie} {et~al.}(2006){Tobie}, {Lunine}, \& {Sotin}}]{Tobie2006}
{Tobie}, G., {Lunine}, J.~I., \& {Sotin}, C. 2006, \nat, 440, 61

\bibitem[{{Triaud} {et~al.}(2013){Triaud}, {Gillon}, {Selsis}, {Winn},
  {Demory}, {Artigau}, {Laughlin}, {Seager}, {Helling}, {Mayor}, {Albert},
  {Anderson}, {Bolmont}, {Doyon}, {Forveille}, {Hagelberg}, {Leconte}, {Lendl},
  {Littlefair}, {Raymond}, \& {Sahlmann}}]{Triaud2013}
{Triaud}, A.~H.~M.~J., {Gillon}, M., {Selsis}, F., {et~al.} 2013, ArXiv
  e-prints

\bibitem[{{von Paris} {et~al.}(2013){von Paris}, {Selsis}, {Kitzmann}, \&
  {Rauer}}]{vonParis2013}
{von Paris}, P., {Selsis}, F., {Kitzmann}, D., \& {Rauer}, H. 2013,
  Astrobiology, 13, 899

\bibitem[{{Williams} \& {Kasting}(1996)}]{Williams1996Kasting}
{Williams}, D.~M. \& {Kasting}, J.~F. 1996, in Lunar and Planetary Science
  Conference, Vol.~27, Lunar and Planetary Science Conference, 1437

\bibitem[{{Williams} \& {Kasting}(1997)}]{WilliamsKasting1997}
{Williams}, D.~M. \& {Kasting}, J.~F. 1997, \icarus, 129, 254

\bibitem[{{Williams} \& {Pollard}(2002)}]{WilliamsPollard2002a}
{Williams}, D.~M. \& {Pollard}, D. 2002, International Journal of Astrobiology,
  1, 61

\bibitem[{{Wong} {et~al.}(2015){Wong}, {Knutson}, {Lewis}, {Kataria},
  {Burrows}, {Fortney}, {Schwartz}, {Agol}, {Cowan}, {Deming}, {D{\'e}sert},
  {Fulton}, {Howard}, {Langton}, {Laughlin}, {Showman}, \&
  {Todorov}}]{Wong2015}
{Wong}, I., {Knutson}, H.~A., {Lewis}, N.~K., {et~al.} 2015, \apj, 811, 122

\bibitem[{{Wordsworth} {et~al.}(2013){Wordsworth}, {Forget}, {Millour}, {Head},
  {Madeleine}, \& {Charnay}}]{Wordsworth2013}
{Wordsworth}, R., {Forget}, F., {Millour}, E., {et~al.} 2013, \icarus, 222, 1

\bibitem[{{Wordsworth} {et~al.}(2010){Wordsworth}, {Forget}, {Selsis},
  {Madeleine}, {Millour}, \& {Eymet}}]{Wordsworth2010}
{Wordsworth}, R.~D., {Forget}, F., {Selsis}, F., {et~al.} 2010, \aap, 522, A22

\bibitem[{{Wordsworth} {et~al.}(2011){Wordsworth}, {Forget}, {Selsis},
  {Millour}, {Charnay}, \& {Madeleine}}]{Wordsworth2011}
{Wordsworth}, R.~D., {Forget}, F., {Selsis}, F., {et~al.} 2011, ApJ Lett., 733,
  L48+

\bibitem[{{Yang} {et~al.}(2013){Yang}, {Cowan}, \& {Abbot}}]{Yang2013}
{Yang}, J., {Cowan}, N.~B., \& {Abbot}, D.~S. 2013, \apjl, 771, L45

\bibitem[{{Yang} {et~al.}(2014){Yang}, {Liu}, {Hu}, \& {Abbot}}]{Yang2014}
{Yang}, J., {Liu}, Y., {Hu}, Y., \& {Abbot}, D.~S. 2014, \apjl, 796, L22

\end{thebibliography}
\end{document}